\documentclass[fleqn,usenatbib]{mnras}

\usepackage{newtxtext,newtxmath}

\usepackage[T1]{fontenc}
\usepackage{ae,aecompl}

\usepackage{multirow}
\usepackage{longtable}

\usepackage{graphicx}	
\usepackage{amsmath}	
\usepackage{amssymb}	

\newcommand{\angstrom}{\mbox{\normalfont\AA}}

\title[H$\alpha$ Morphologies of Star Clusters]{H$\alpha$ Morphologies of Star Clusters: A LEGUS study of HII region evolution timescales and stochasticity in low mass clusters}
\author[Hannon S. et al.]{Stephen Hannon,$^{1,2,3}$\thanks{E-mail: shann004@ucr.edu}
Janice C. Lee,$^{2,3}$
B.C. Whitmore,$^{4}$
R. Chandar,$^{5}$
A. Adamo,$^{6}$
\newauthor
B. Mobasher,$^{1}$
A. Aloisi,$^{4}$
D. Calzetti,$^{7}$
M. Cignoni,$^{8}$
D.O. Cook,$^{2,3}$
D. Dale,$^{9}$
S. Deger,$^{3}$
\newauthor
L. Della Bruna,$^{6}$
D.M. Elmegreen,$^{10}$
D.A. Gouliermis,$^{11,12}$
K. Grasha,$^{7}$
E.K. Grebel,$^{13}$
\newauthor
A. Herrero,$^{14,15}$
D.A. Hunter,$^{16}$
K.E. Johnson,$^{17}$
R. Kennicutt,$^{18,19}$
H. Kim,$^{20}$
\newauthor
E. Sacchi,$^{4}$
L. Smith,$^{4}$
D. Thilker,$^{21}$
J. Turner,$^{9}$
R.A.M. Walterbos,$^{22}$
A. Wofford$^{23}$
\\
$^{1}$Department of Physics $\And$ Astronomy, University of California, Riverside, CA, USA\\
$^{2}$Department of Physics $\And$ Astronomy, California Institute of Technology, Pasadena, CA, USA\\
$^{3}$IPAC, California Institute of Technology, Pasadena, CA, USA\\
$^{4}$Space Telescope Science Institute, Baltimore, MD, USA\\
$^{5}$Department of Physics and Astronomy, University of Toledo, Toledo, OH, USA\\
$^{6}$Department of Astronomy, The Oskar Klein Centre, Stockholm University, Stockholm, Sweden\\
$^{7}$Department of Astronomy, University of Massachusetts, Amherst, MA 01003, USA\\
$^{8}$Department of Physics, University of Pisa, Largo B. Pontecorvo 3, 56127, Pisa, Italy\\
$^{9}$Department of Astronomy, University of Wyoming, Laramie, WY\\
$^{10}$Department of Physics and Astronomy, Vassar College, Poughkeepsie, NY\\
$^{11}$Zentrum f{\"u}r Astronomie der Universit{\"a}t Heidelberg, Institut f{\"u}r Theoretische Astrophysik, Albert-Ueberle-Str. 2, Heidelberg, Germany\\
$^{12}$Max Planck Institute for Astronomy, K{\"o}nigstuhl 17, 69117 Heidelberg, Germany\\
$^{13}$Astronomisches Rechen-Institut, Zentrum f{\"u}r Astronomie der Universit{\"a}t Heidelberg, M{\"o}nchhofstr. 12-14, 69120 Heidelberg, Germany\\
$^{14}$Instituto de Astrofisica de Canarias, La Laguna, Tenerife, Spain\\
$^{15}$Departamento de Astrofisica, Universidad de La Laguna, Tenerife, Spain\\
$^{16}$Lowell Observatory, Flagstaff, AZ\\
$^{17}$Department of Astronomy, University of Virginia, Charlottesville, VA\\
$^{18}$Institute of Astronomy, University of Cambridge, Cambridge, United Kingdom\\
$^{19}$Department of Astronomy, University of Arizona, Tucson, AZ\\
$^{20}$Gemini Observatory, Casilla 603, La Serena, Chile\\
$^{21}$Department of Physics and Astronomy, The Johns Hopkins University, Baltimore, MD\\
$^{22}$Department of Astronomy, New Mexico State University, Las Cruces, NM\\
$^{23}$Instituto de Astronomia, Universidad Nacional Autonoma de Mexico, Unidad Academica en Ensenada, Km 103 Carr., Ensenada, Mexico
}

\date{Accepted 2019 October 03. Received 2019 September 14; in original form 2019 June 03}

\pubyear{2019}

\begin{document}
\label{firstpage}
\pagerange{\pageref{firstpage}--\pageref{lastpage}}
\maketitle

\begin{abstract}
The morphology of HII regions around young star clusters provides insight into the timescales and physical processes that clear a cluster's natal gas.  We study $\sim$700 young clusters ($\leq$10Myr) in three nearby spiral galaxies (NGC 7793, NGC 4395, and NGC 1313) using Hubble Space Telescope (\textit{HST}) imaging from LEGUS (Legacy ExtraGalactic Ultraviolet Survey).  Clusters are classified by their H$\alpha$ morphology (concentrated, partially exposed, no-emission) and whether they have neighboring clusters (which could affect the clearing timescales). Through visual inspection of the \textit{HST} images, and analysis of ages, reddenings, and stellar masses from spectral energy distributions fitting, together with the (U-B), (V-I) colors, we find:  1) the median ages indicate a progression from concentrated ($\sim$3 Myr), to partially exposed ($\sim$4 Myr), to no H$\alpha$ emission (>5Myr), consistent with the expected temporal evolution of HII regions and previous results. However, 2) similarities in the age distributions for clusters with concentrated and partially exposed H$\alpha$ morphologies imply a short timescale for gas clearing ($\lesssim$1Myr). Also, 3) our cluster sample's median mass is $\sim$1000 $M_\odot$, and a significant fraction ($\sim$20$\%$) contain one or more bright red sources (presumably supergiants), which can mimic reddening effects.  Finally, 4) the median E(B-V) values for clusters with concentrated H$\alpha$ and those without H$\alpha$ emission appear to be more similar than expected ($\sim$0.18 vs. $\sim$0.14, respectively), but when accounting for stochastic effects, clusters without H$\alpha$ emission are less reddened.  To mitigate stochastic effects, we experiment with synthesizing more massive clusters by stacking fluxes of clusters within each H$\alpha$ morphological class. Composite isolated clusters also reveal a color and age progression for H$\alpha$ morphological classes, consistent with analysis of the individual clusters.

\end{abstract}

\begin{keywords}
ISM: HII regions -- ISM: evolution -- galaxies: star clusters -- galaxies: ISM
\end{keywords}



\section{Introduction}
\label{sec:section1}

The study of young star clusters is crucial for understanding the formation and evolution of stars in general, as most of the star formation in our universe occurs in a clustered fashion \citep{LADA03}. Star clusters are born in clouds of cold gas which are subsequently ionized by massive OB-type stars formed within the clusters, resulting in nebular H$\alpha$ emission.

Stellar feedback within these star-forming regions significantly affects the size, shape and extent of this H$\alpha$ emission 
(e.g. \citealt{KENNICUTT84,CHURCHWELL06,WHITMORE11,ANDERSON14}). By examining the morphology of these HII regions across a large cluster sample, we can investigate questions such as: 1) Is there an evolutionary pattern found amongst different morphological types? 2) If so, what is the timescale associated with the evolution? 3) What physical processes drive the evolution? 4) What other properties are correlated with HII region morphology and why?

Previous studies have examined correlations between H$\alpha$ morphology and star cluster properties, derived from the fitting of Spectral Energy Distributions (SEDs) to \textit{HST} photometry. For example, \citet{WHITMORE11} investigated the use of H$\alpha$ morphologies of the star cluster population in M83 to constrain cluster ages.  This led to a classification scheme for H$\alpha$ morphologies representing different stages in a cluster's evolution (from concentrated H$\alpha$ to a gradually more blown-out region). Whitmore et al. argue that this classification scheme provides a viable method for dating young ($\sim$1-10Myr) clusters.  \citet{HOLLYHEAD15} also examined the young ($\leq$ 10Myr), massive ($\geq$5000$M_\odot$) clusters of M83 by including data from 6 additional \textit{HST} fields. From the H$\alpha$ morphologies, they inferred how much time clusters spent in an "embedded" state, thereby providing insight into the timescales associated with gas clearing for their sample. In addition to finding an evolutionary pattern generally consistent with the results of \citet{WHITMORE11}, they concluded that the gas removal process began in most clusters at ages of 2-3 Myr, implying that supernovae (SNe) could not be the sole driver. Using a different approach by comparing ALMA CO detections of giant molecular clouds (GMCs) with the positions of star clusters in NGC 7793, \citet{GRASHA18} found that the timescale for star clusters to dissociate from their natal clouds is similarly very short, between 2-3Myr, roughly consistent with the timescales found based on ALMA CO data for the Antennae galaxies \citep{MATTHEWS18}.

Here, we study the H$\alpha$ morphologies of the ionized gas surrounding star clusters in a sample of three nearby spiral galaxies (NGC 1313, NGC 4395, NGC 7793) and retain the more common low-mass clusters in our analysis.  Observed star cluster mass functions appear to be well-fit by power laws of the form dN/dM $\propto$ M$^{-2}$ (e.g., \citealt{WILLIAMS97,ZHANG99,LARSEN02,BASTIAN11,FALL12,FOUESNEAU12}), implying that clusters in the mass range 10$^{2.5}$-10$^{3.5}$ $M_\odot$ are nearly ten times as numerous as those in the range 10$^{3.5}$-10$^{5.5}$ $M_\odot$. However, analyses of lower-mass clusters have been shown to be affected by the stochastic sampling of the upper part of the stellar mass function (e.g. \citealt{BARBARO77,GIRARDI93,LANCON00,BRUZUAL02,CERVINO06,DEVEIKIS08}), making it challenging to study their physical properties.  In this low-mass regime, the predicted luminosity and color distributions can be far from Gaussian even when the total cluster mass exceeds 10$^{5}$ $M_\odot$, as they depend strongly on the mass distribution of stars in the cluster (\citealt{FOUESNEAU10,FOUESNEAU12}). 

This paper examines these stochastic sampling effects observationally, and aims to address the following questions: 1) Is there evidence for stochastic effects in the properties of clusters (color, age, mass, reddening)? 2) What potential methods could be employed to mitigate against stochastic effects and provide better constraints on cluster properties? Addressing these questions will provide important insight into the nature of star clusters, and possible strategies for future studies of low-mass star clusters, which constitute the bulk of the population.

In Section~\ref{sec:section2} we summarize the observations and the star cluster catalogs used in the study. Section~\ref{sec:section3} describes the visual classification scheme used for H$\alpha$ morphologies. Section~\ref{sec:section4} examines the cluster age and reddening distributions as a function of H$\alpha$ morphological class. In Section~\ref{sec:section5}, we investigate the effects of stochastic sampling of the stellar IMF and examine methods to mitigate against its impact. In Section~\ref{sec:section6}, we discuss the implications of our results for H$\alpha$ morphology evolution and compare our study with previous works. Section~\ref{sec:section7} provides an overall summary of the work as well as potential future studies.

\section{Data}
\label{sec:section2}

\subsection{Observations}

We use data from the Legacy ExtraGalactic UV Survey (LEGUS; \citealt{CALZETTI15}), which has collected \textit{HST} imaging with the Wide Field Camera 3 (F275W, F336W, F438W, F555W, and F814W) for 50 nearby (within $\sim$12 Mpc) galaxies spanning a range of properties. WFC3 observations taken specifically for the LEGUS program in Cycle 21 (GO-13364) are combined with ACS data taken in previous cycles by other programs to provide full 5-band coverage for the LEGUS sample. We also use data from the LEGUS-H$\alpha$ follow-up survey (GO-13773; PI R. Chandar), where a narrow-band filter covering the H$\alpha$ emission-line (F657N) and a medium-band filter sampling the line-free continuum (F547M) were used to image the 25 LEGUS galaxies with the highest star formation rates.  To produce H$\alpha$ emission-line images\footnote{Note that the emission-line images also contain flux from the adjacent [NII] 6548,83 line.}, the drizzled and aligned F657N images are continuum subtracted using an image formed from a combination of F814W and F547M, appropriately scaled using their AB zeropoints.

\begin{figure}
\includegraphics[width =0.45\textwidth]{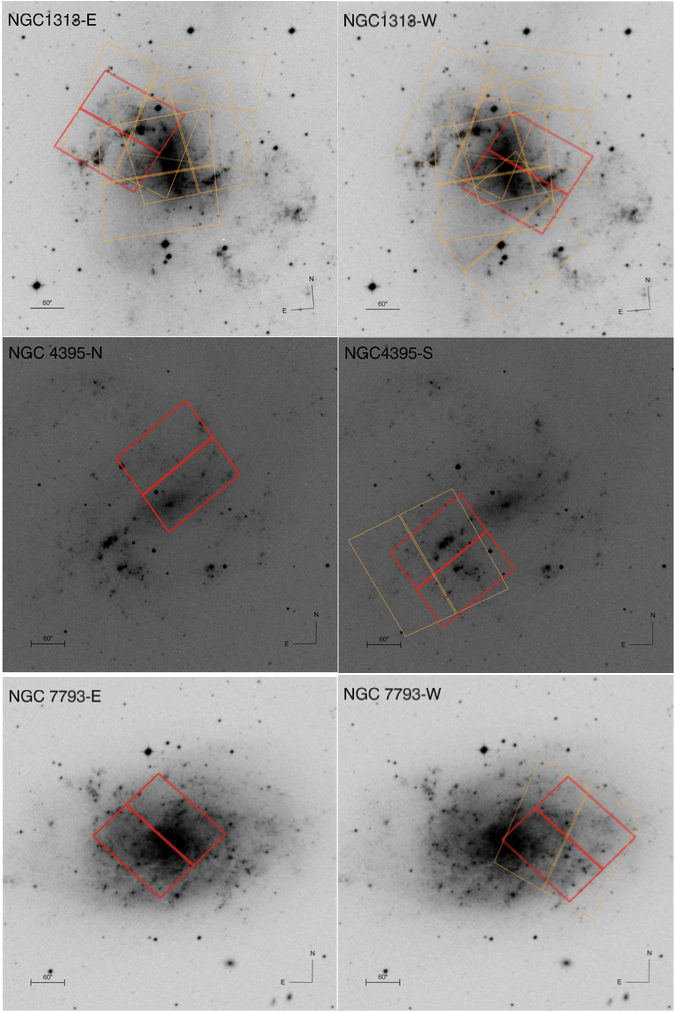}
\caption{\textit{HST} footprints on digitized sky survey (DSS) images for each of the 6 fields used in this study: NGC 1313-E, NGC 1313-W, NGC 4395-N, NGC 4395-S, NGC 7793-W, and NGC 7793-E. Red outlines represent WFC3 images and orange outlines, where available, represent previously available ACS images. The field of view of WFC3 is 162" x 162".}
\label{fig:Figure1}
\end{figure}

For this study, we have selected the three nearest (d $\approx$ 4Mpc) spiral galaxies (six \textit{HST} pointings) from the LEGUS survey: NGC 1313 (E and W pointings), NGC 4395 (N and S pointings), and NGC 7793 (E and W pointings), all of which have \textit{HST} H$\alpha$ emission-line and continuum imaging. We choose the three nearest galaxies to maximize resolution for H$\alpha$ morphological classification. Fig.~\ref{fig:Figure1} shows the \textit{HST} footprints for each of the pointings for the three galaxies in this study.

Details on filters, exposure times, and other pertinent information on selection techniques and properties of these galaxies can be found in \citet{CALZETTI15}. 

\subsection{Star Cluster Catalogs}
\label{sec:Star Cluster Catalogs}

Catalogs of the photometric and physical properties of star clusters for the majority of LEGUS galaxies have been developed and publicly released by the LEGUS team (https://legus.stsci.edu)\footnote{Catalogs for NGC 4395 are based on updated models and will be posted in a future release.}. Here, we provide a short overview of the selection criteria, as we use these catalogs to select young star clusters for our analysis. A full description of the selection methods, photometry, and derivation of cluster masses, ages, and reddening via SED fitting are given in \citet{ADAMO17} and \citet{COOK19}.

The star clusters in LEGUS were selected using the following criteria:
\begin{enumerate}
    \item detection in at least 4 filters (UBVI, UV-BVI, or UV-UBVI) with photometric error less than 0.3 mag.
    \item $M_V$ $\leq$ -6.0 (using the F555W filter)
    \item a concentration index (CI) threshold, where CI = mag(1pix) - mag(3pix) in F555W to separate point sources from extended sources.  For the three galaxies here, the CI threshold is between $\geq$ 1.2 and 1.4, depending on distance of the galaxy and the camera used to obtain the F555W imaging (i.e., ACS or WFC3).
\end{enumerate}

All candidates are visually inspected by three LEGUS team members, and placed into the following classes:

\vspace{2mm}

Class = 1: Symmetric, compact cluster

Class = 2: Concentrated object with some degree of asymmetry; possible color gradient

Class = 3: Multiple peak system; compact association; could be confused with spurious sources if there are nearby stars along the line of sight

Class = 4: Spurious detection (e.g. foreground/background sources, single bright stars, artifacts)

\vspace{2mm}


Photometry is performed with apertures with radii of 4, 5, and 6 pixels for clusters in NGC 4395, NGC 7793, and NGC 1313, respectively. 
At the adopted distances of 4.30, 3.44, and 4.39 Mpc \citep{CALZETTI15}, these apertures subtend 3.3, 3.3, and 5.1 pc, respectively, given a WFC3 UVIS pixel scale of .04 arcseconds per pixel.  
To measure and subtract the background, a 1 pixel-wide sky annulus at a 7 pixel radius is used.  An average aperture correction is estimated as the difference between the magnitude of the source measured at 20 pixels (with a 1 pixel-wide sky annulus at a 21 pixel radius) minus the magnitude of the source obtained using the smaller (4, 5, and 6 pixel) aperture for a set of bright isolated clusters.  The total cluster magnitudes are corrected for foreground Galactic extinction \citep{SCHLAFLY11}.

Cluster SEDs are fitted based on the \textit{HST} photometry with Yggdrasil SSP models \citep{ZACKRISSON11}, which include nebular flux via photoionization modeling with CLOUDY, assuming a covering fraction of 0.5.\footnote{To validate the use of the models which include nebular emission to infer the ensemble properties of our star cluster sample, H$\alpha$ equivalent widths (EWs) are measured for isolated clusters with concentrated H$\alpha$ morphologies. For ages less than 4 Myr, the model values of log(EW[H$\alpha$] ({\angstrom})) range from $\sim$2.7-3.25. The measured values are consistent within the measurement uncertainties: log(EW[H$\alpha$] ({\angstrom})) ranges from 2.6-3.6 with a median of 3.1. More detailed model validation is advised if examining the properties of individual objects, as opposed to the ensemble properties of interest here.}  The model grid used in the fitting is based on 46 time progressive steps from 1Myr to 10Gyr, and 150 fixed steps in reddening from 0.00 to 1.50. The "best values" used in this work are those corresponding to the minimum $\chi^{2}$ value. 
For each galaxy, the present-day metallicity of its young populations as derived from nebular abundances (listed in \citealt{CALZETTI15}) is adopted. This corresponds to a metallicity of Z = 0.02 (solar) for clusters in NGC 7793 and NGC 1313 while those in NGC 4395 assume a metallicity of Z = 0.004. 

The errors of the ages, masses, and reddenings used in this study are based on the minimum and maximum values of the SED fitted parameters. Based on 1$\sigma$ confidence levels of Lampton et al. (1976), all age and reddening values with corresponding $\chi^{2}$ $\leq$ $\chi^{2}_{min}$ + 2.3 are found, and from these values, minima and maxima are determined. In the cases where no other grid solution in the SED fitting procedure are found, the errors are half the size of the age step (1 Myr) or reddening step (0.01). 

A total of twelve catalogs have been produced for each galaxy based on two stellar evolution models (Geneva and Padova), three extinction models (Milky Way extinction, starburst extinction, and differential starburst extinction), and two aperture correction methods (concentration index based and average aperture correction). 
In this work, we use the LEGUS "reference" catalogs \citep{ADAMO17}, based on the Padova stellar evolution models, Milky Way extinction, and using standard average aperture correction, but it should be noted that the Geneva models are generally considered to better describe the youngest population of stars, and differences in the results of our analysis are discussed later in this paper. It should also be noted that the SED fitting results used in this analysis are only based on the broadband photometry and are independent of the narrow and medium band photometry. Information from the H$\alpha$ imaging only enters the analysis once, in the classification of the cluster H$\alpha$ morphologies. 

For our study, we produce a sample of verified clusters by selecting all objects with mode class of 1, 2, 3 and excluding sources whose H$\alpha$ images were partially cut off by the edge of the field of view. We then narrow our sample down to those with best-fitted ages of 10 Myr or younger, leaving us with a total of 654 final cluster targets across all 6 pointings. Table~\ref{tab:table1} shows the total number of objects in the overall catalogs, as well as in our young cluster sample, for the galaxies in this study.

\begin{table*}
\begin{tabular}{lllll}
\hline
\bfseries Galaxy & \bfseries Total Detections & \bfseries Verified Clusters & \bfseries $\leq$ 10Myr & \bfseries $\geq$ 5000 $M_\odot$ \\
\hline
NGC1313W & 3784  & 486      & 177   & 1       \\
NGC1313E & 5284  & 259      & 111   & 2       \\
NGC4395N & 291   & 39       & 23    & 14      \\
NGC4395S & 837   & 137      & 112   & 18      \\
NGC7793W & 2794  & 221      & 119   & 0       \\
NGC7793E & 899   & 191      & 112   & 7       \\
\hline
Totals   & 13889 & 1333     & 654   & 42 \\
\hline
\end{tabular}
\caption{Number counts of total detections, visually-verified clusters (LEGUS mode class = 1, 2, or 3), verified clusters younger than 10 Myr, and verified clusters greater than 5000 $M_\odot$ for each of the six fields of view. Masses and ages are determined by comparing the measured broad-band photometry with the Yggdrasil models, and assuming Milky Way extinction, Padova stellar evolutionary tracks for an appropriate metallicity, and the average aperture correction procedure described in \citet{ADAMO17}}
\label{tab:table1}
\end{table*}

\section{H\texorpdfstring{$\alpha$}{} Morphology Classification}
\label{sec:section3}

To aid our visual classification process, two sets of 150pc x 150pc postage stamps were created for each cluster. One set of stamps was made from an RGB image of the galaxy using combined NUV and U bands for the blue channel, combined V and I bands for the green channel, and the continuum-subtracted H$\alpha$ narrow band for the red channel. The second set of postage stamps was created solely from the continuum-subtracted H$\alpha$ narrow band image. Whereas the narrow band images clearly display the shape and extent of the HII regions, the RGB images are useful in showing the target cluster and its neighbors in relation to the surrounding gas. 

\begin{figure}
\includegraphics[width =0.5\textwidth]{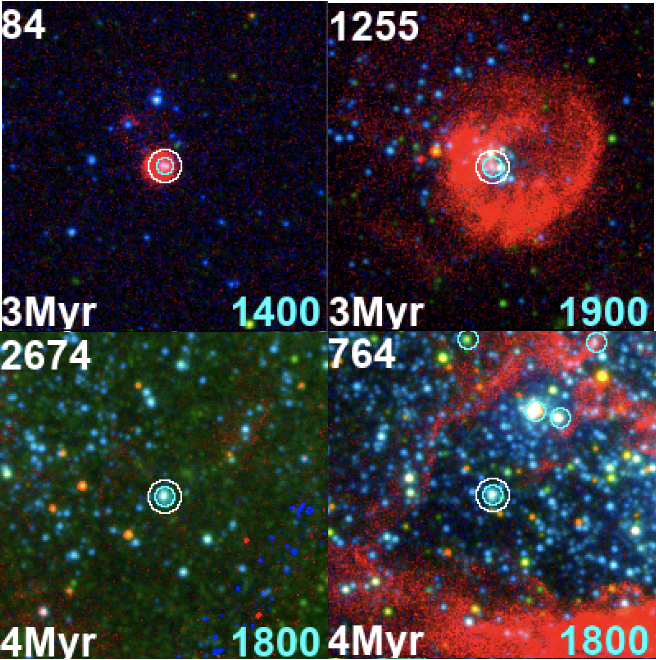}
\caption{RGB postage stamps exemplifying the three HII morphology classifications. The red in the images corresponds to continuum-subtracted H$\alpha$. In each stamp, the object ID is located in the upper left, age in the lower-left, and mass (in $M_\odot$) in the lower-right. White circles with radii of 7.5pc are drawn around the stamp's target cluster. Cyan circles represent the photometric aperture radii ($\sim$4pc). Clockwise from the upper-left panel: Object 84 from NGC 4395-N has an estimated age of 3 Myr and is classified as concentrated. Object 1255 from NGC 1313-E is also estimated to be 3 Myr and classified as partially exposed. Object 2674 from NGC 1313-E is $\sim$4 Myr and is classified as no-emission. Object 764 from NGC 1313-E is also $\sim$4 Myr and also classified as no-emission and is the only non-isolated cluster shown here. All postage stamps are oriented north-up, east-left.}
\label{fig:Figure2}
\end{figure}

As illustrated in Fig.~\ref{fig:Figure2}, each individual stamp includes small cyan circles to indicate the positions of all clusters (including those older than 10 Myr). The radii of these circles represent the aperture sizes used for photometry in that particular field; i.e., whereas apertures with 6-pixel radii were used for sources in NGC 1313, 4-pixel and 5-pixel radii apertures were used for NGC 4395 and NGC 7793, respectively (see Section \ref{sec:Star Cluster Catalogs}).  These aperture sizes roughly correspond to a physical radius of 4pc. In addition, a white circle of radius 7.5pc is drawn around the central target as a reference to help determine the compactness of the HII region. 

For reference, an O5 star producing $\sim$5 x 10$^{49}$ Lyman continuum photons per second\footnote{\citet{OSTERBROCK06}} would create an idealized Str\"omgren sphere with a radius between $\sim$1.1pc and 110pc for gas densities between $n_e$ $\approx$ $n_p$ = 1 cm$^{-3}$ (larger sphere) and $n_e$ $\approx$ $n_p$ = 10$^{3}$ cm$^{-3}$ (smaller sphere). In the same range of gas densities, a B0 star producing $\sim$5 x 10$^{47}$ ionizing photons per second would create a Str\"omgren radius between $\sim$0.25pc and 25pc. These numbers assume a uniform ISM rather than a more realistic clumpy ISM, but are used here as a guide.


Our morphological classification scheme is based on one used by \citet{HOLLYHEAD15}, who defined three classes based on the presence and shape of the H$\alpha$ emission, similar to that of \citet{WHITMORE11}:

\renewcommand{\labelenumi}{\arabic{enumi}}
\begin{enumerate}
    \item \textbf{concentrated}, where the target star cluster has a compact HII region and where there are no discernable bubbles or areas around the cluster which lack H$\alpha$ emission,
    \item \textbf{partially exposed}, where the target cluster shows bubble like/filamentary morphology covering part of the cluster and, 
    \item \textbf{no emission}, where the target cluster does not appear to be associated with H$\alpha$. There is no H$\alpha$ emission within $\sim$20pc of the cluster. The majority (62$\%$) are clusters without any visible H$\alpha$ emission in their 150 parsec-wide postage stamps.
\end{enumerate}

An important consideration in classifying clusters by their H$\alpha$ morphology is the detection limit for H$\alpha$ and the stars required to produce the ionizing radiation. The most straightforward way to characterize the sensitivity of the H$\alpha$ images is to compute the total flux of a point source at a specified detection limit.  For the H$\alpha$ images used in this study, we find we that the 5$\sigma$ point source detection limit is between 5.0x10$^{-17}$ and 5.5x10$^{-17}$ ergs cm$^{-2}$ s$^{-1}$.  At a distance of 4 Mpc, this gives an observed (i.e. no extinction corrections applied) luminosity of $\sim$1.5x10$^{35}$ erg $^{-1}$. Given the model grid of \citet{SMITH02}, this luminosity corresponds to the ionizing flux of a B0.5V star ($\sim$10$M_\odot$), with the usual assumptions: solar metallicity, Case B recombination, nebular temperatures and densities of 10$^{4}$ K and 100 cm$^{-3}$ respectively, and that the nebular region is radiation bounded. 

In addition to these three H$\alpha$ morphological classes, each cluster is also identified as either isolated or non-isolated. \citet{GOULIERMIS18} compiled survey data for several thousand unbound stellar systems in Local Group galaxies and found that the average size of these stellar associations is $\sim$70-90pc. Comparably, we define an isolated cluster to be one that does not have any neighboring clusters within 75pc. This is done to control for the potential impact of neighboring clusters on the clearing of a target cluster's gas, which may have a confounding effect on correlations between cluster properties and the H$\alpha$ morphology. 

Our hypothesis in classifying the H$\alpha$ morphology of the young clusters in such a manner is that they represent HII region evolutionary stages.  We are interested in testing for possible correlations between the H$\alpha$ morphology and cluster age, reddening, and mass. Fig.~\ref{fig:Figure2} shows RGB postage stamps of examples from each of the three H$\alpha$ morphological classes, with the cluster ID in a given pointing labeled in the upper left-hand corner, its estimated age (in Myr) in the lower left-hand corner, and its estimated mass (in $M_\odot$) in the bottom right-hand corner. The isolated object in the upper-left panel (ID 84) is classified as concentrated, as the HII region has a size comparable to the 7.5pc radius circular marker and is fairly compact. The upper-right panel shows an isolated cluster with H$\alpha$ morphology classified as partially exposed (ID 1255), as there is little emission at the center of the HII region at the position of the cluster, creating an apparent bubble. The lower-left panel shows an isolated cluster without H$\alpha$ emission (ID 2674), as there is no HII region clearly associated with the cluster. Finally, the lower-right panel also shows a cluster with morphology classified as no emission (ID 764), as the H$\alpha$ is tens of parsecs away in every direction and it is not clear whether the cluster is contributing ionizing radiation responsible for the H$\alpha$ emission; this is also an example of a non-isolated cluster as there are several other clusters in the vicinity. We note that while our H$\alpha$ classification does not distinguish between the two no-emission examples in Fig.~\ref{fig:Figure2}, the majority (>60$\%$) of isolated clusters in this class showed no clear HII region associated with the cluster, resembling object ID 2674.

\begin{table*}
\begin{tabular}{lllllll}
\hline
                      & \multicolumn{2}{c}{Concentrated} & \multicolumn{2}{c}{Partially Exposed} & \multicolumn{2}{c}{No Emission} \\
                      \hline
Galaxy                & Isolated       & Non-Isolated    & Isolated         & Non-Isolated       & Isolated       & Non-Isolated   \\
\hline
NGC1313W              & 5              & 25              & 1                & 20                 & 39             & 87             \\
NGC1313E              & 10             & 15              & 1                & 25                 & 29             & 31             \\
NGC4395N              & 6              & 0               & 1                & 0                  & 14             & 2              \\
NGC4395S              & 9              & 16              & 4                & 23                 & 20             & 40             \\
NGC7793W              & 10             & 20              & 3                & 19                 & 37             & 30             \\
NGC7793E              & 11             & 15              & 4                & 11                 & 47             & 24             \\
\hline
\bfseries Totals                & \bfseries 51 (7.8\%)     & \bfseries 91 (13.9\%)     & \bfseries 14 (2.1\%)       & \bfseries 98 (15.0\%)       & \bfseries 186 (28.4\%)   & \bfseries 214 (32.7\%)   \\
$\geq$ 5000 $M_\odot$ & 0              & 5               & 0                & 7                  & 13             & 17 \\
\hline
\end{tabular}
\caption{Number counts of clusters in each of the three morphology classes (concentrated, partially exposed, no-emission) and further separated into isolated and non-isolated categories depending on the presence of neighboring clusters within 75pc. The total number of clusters (and their percentage of the total 654) and the number of clusters greater than or equal to 5000 $M_\odot$ in each of the six bins are provided in the bottom two rows.}
\label{tab:table2}
\end{table*}

The number of clusters in each H$\alpha$ class and isolation category is shown in Table~\ref{tab:table2}.  The majority of clusters are classified as no-emission ($\sim$60$\%$).   Although the fractions of isolated and non-isolated clusters are comparable for the concentrated and no-emission classes, it is noteworthy that the partially exposed clusters of our sample show a much greater ratio of non-isolated to isolated clusters ($\sim$7:1).   This possible reasons for this can be further examined in future work, upon expansion of the dataset.

\section{Results}
\label{sec:section4}

In this section, we examine the ages, masses and reddening of clusters in each morphological class and isolation category based on both the Padova and Geneva stellar evolutionary models. The mean and median age, reddening, and mass for each of these categories are summarized in Table~\ref{tab:table3}. The following subsections detail the characteristics of the age and reddening distributions.

\begin{table*}
\begin{tabular}{llllllllll}
\hline
             &           & \multicolumn{4}{c}{Padova}                                       & \multicolumn{4}{c}{Geneva}                                       \\
             \hline
H$\alpha$ Class     & Isolated? & N   & Mean \textbf{Age} (Myr)        & Median \textbf{Age} (Myr)          & SD  & N   & Mean \textbf{Age} (Myr)        & Median \textbf{Age} (Myr)          & SD  \\
\hline
Concentrated & Iso      & 51  & 3.3                   & 3.0 {[}2.0, 4.0{]}        & 1.6    & 51  & 2.6                   & 3.0 {[}2.0, 3.0{]}        & 1.0    \\
             & Non-Iso     & 91  & 3.1                   & 3.0 {[}1.0, 4.0{]}        & 1.7    & 91  & 2.7                   & 3.0 {[}2.0, 3.0{]}        & 1.4    \\
Partially    & Iso      & 14  & 3.9                   & 4.0 {[}2.3, 5.0{]}        & 2.1    & 14  & 3.0                   & 3.0 {[}3.0, 3.0{]}        & 0.9    \\
Exposed      & Non-Iso     & 98 & 4.1                   & 4.0 {[}3.0, 5.0{]}        & 2.1    & 98  & 3.2                   & 3.0 {[}3.0, 3.0{]}        & 1.3    \\
No Emission  & Iso      & 186 & 5.8                   & 5.0 {[}4.0, 8.0{]}        & 2.5    & 201 & 5.1                   & 5.0 {[}3.0, 6.0{]}        & 2.2    \\
             & Non-Iso     & 214 & 4.7                   & 4.0 {[}3.0, 6.0{]}        & 2.4    & 229 & 4.0                   & 3.0 {[}3.0, 5.0{]}        & 2.0    \\
             \hline
             \hline
H$\alpha$ Class     & Isolated? & N   & Mean \textbf{E(B-V)}        & Median \textbf{E(B-V)}          & SD  & N   & Mean \textbf{E(B-V)}        & Median \textbf{E(B-V)}          & SD  \\
\hline
Concentrated & Iso      & 51  & 0.19                  & 0.17 {[}0.08, 0.28{]}     & 0.16   & 51  & 0.21                  & 0.19 {[}0.08, 0.29{]}     & 0.16   \\
             & Non-Iso     & 91  & 0.18                  & 0.13 {[}0.05, 0.26{]}     & 0.15   & 91  & 0.17                  & 0.14 {[}0.05, 0.27{]}     & 0.14   \\
Partially    & Iso      & 14  & 0.10                  & 0.07 {[}0.01, 0.15{]}     & 0.11   & 14  & 0.10                  & 0.06 {[}0.01, 0.11{]}     & 0.14   \\
Exposed      & Non-Iso     & 98 & 0.08                  & 0.05 {[}0.00, 0.11{]}     & 0.13   & 98  & 0.09                  & 0.04 {[}0.00, 0.10{]}     & 0.19   \\
No Emission  & Iso      & 186 & 0.20                  & 0.14 {[}0.05, 0.25{]}     & 0.22   & 201 & 0.17                  & 0.13 {[}0.05, 0.23{]}     & 0.18   \\
             & Non-Iso     & 214 & 0.14                  & 0.08 {[}0.02, 0.18{]}     & 0.19   & 229 & 0.16                  & 0.09 {[}0.03, 0.21{]}     & 0.22   \\
             \hline
             \hline
H$\alpha$ Class     & Isolated? & N   & Mean \textbf{Mass}($M_\odot$) & Median \textbf{Mass} ($M_\odot$)   & SD  & N   & Mean \textbf{Mass}($M_\odot$) & Median \textbf{Mass} ($M_\odot$)   & SD  \\
\hline
Concentrated & Iso      & 51  & 1.1 E3                & 9.0 E2 {[}4.0E2, 1.5E3{]} & 900 & 51  & 1.0 E3                & 7.0 E2 {[}4.0E2, 1.3E3{]} & 800 \\
             & Non-Iso     & 91  & 2.4 E3                & 1.0 E3 {[}6.0E2, 2.1E3{]} & 5700 & 91  & 2.4 E3                & 9.0 E2 {[}5.0E2, 2.1E3{]} & 6200 \\
Partially    & Iso      & 14  & 8.0 E2                & 6.0 E2 {[}5.0E2, 1.3E3{]} & 500 & 14  & 9.0 E2                & 4.0 E2 {[}3.0E2, 1.1E3{]} & 1500 \\
Exposed      & Non-Iso     & 98 & 1.9 E3                & 1.0 E3 {[}5.0E2, 1.8E3{]} & 3000 & 98  & 2.0 E3                & 9.0 E2 {[}4.0E2, 1.6E3{]} & 4100 \\
No Emission  & Iso      & 186 & 2.2 E3                & 1.0 E3 {[}6.0E2, 2.0E3{]} & 3900 & 201 & 1.4 E3                & 7.0 E2 {[}4.0E2, 1.5E3{]} & 2000 \\
             & Non-Iso     & 214 & 2.5 E3                & 1.2 E3 {[}7.0E2, 2.0E3{]} & 5600 & 229 & 2.8 E3                & 1.0 E3 {[}6.0E2, 1.9E3{]} & 6300 \\
             \hline
\end{tabular}
\caption{Mean and median age, reddening, and mass for the clusters aggregated over all six fields in each of the three morphology classes, isolation category, and stellar population model (Padova vs. Geneva). The standard deviations of each distribution is also provided.}
\label{tab:table3}
\end{table*}

\subsection{Age Statistics by Morphological Class}
\label{sec:section41}

\begin{figure*}
\includegraphics[width =1.0\textwidth]{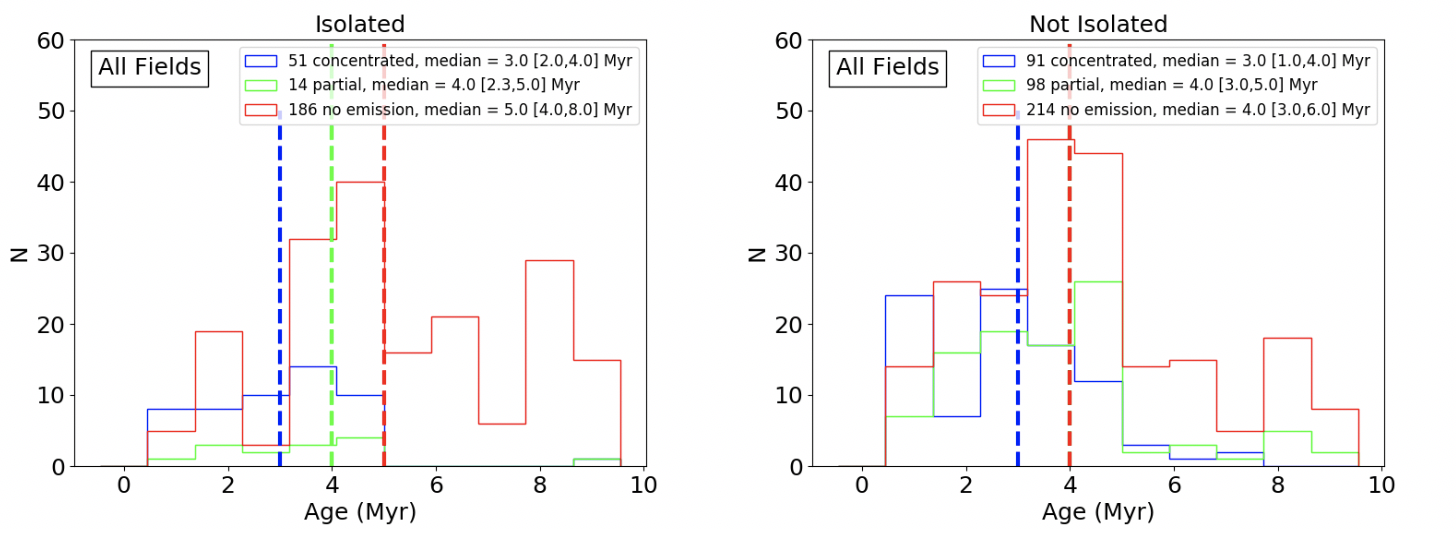}
\centering
\caption{Histograms showing the age distribution of clusters across all the fields in the sample, using the Padova stellar evolutionary model. The left panel shows the 251 isolated clusters in the sample; the right panel shows the 403 clusters with neighboring clusters within 75 pc. The solid blue, green, and red lines represent clusters with H$\alpha$ emission that is concentrated, partially exposed, and absent (i.e., no emission), respectively. The number of clusters as well as the median cluster ages [first quartile, third quartile] for each of the three classes is displayed in the legend. Vertical dashed lines represent the medians for each class.}
\label{fig:Figure3}
\end{figure*}

A key aspect in our analysis of young clusters is examining the age distributions as a function of H$\alpha$ morphological class. These ages can establish whether or not the morphological classes constitute an evolutionary sequence for HII regions. Fig.~\ref{fig:Figure3} shows the age distributions for each of the H$\alpha$ morphological classes based on our reference catalog; the left and right plots show the distributions for the 251 isolated clusters and the 403 non-isolated clusters, respectively.

For the isolated Padova-based cluster sample, we find that the median best age [first quartile, third quartile] of the 51 clusters with concentrated HII regions is 3.0 [2.0, 4.0] Myr, while the ages for the 14 partially exposed clusters and 186 clusters with no H$\alpha$ emission are 4.0 [2.3, 5.0] Myr and 5.0 [4.0, 8.0] Myr, respectively. The non-isolated clusters provide roughly consistent results, with median best ages of 3.0 [1.0, 4.0] Myr, 4.0 [3.0, 5.0] Myr, and 4.0 [3.0, 6.0] Myr for the 91 concentrated, 98 partially exposed, and 214 no-emission H$\alpha$ morphologies, respectively. It is also important to note here that the median ages determined for clusters with no H$\alpha$ emission represent lower limits, due to the initial selection which eliminated clusters with estimated ages older than 10 Myr. While the Padova model shows some differences in the median ages which could possibly indicate an age progression, the Geneva model shows fewer differences, with equal median ages for the concentrated and partially exposed classes in both the isolated and not-isolated samples (3.0 Myr). Regardless of isolation criteria or stellar population model, the mean age of each morphological class does indicate a progression from concentrated to partially exposed to no-emission H$\alpha$ morphologies as expected, though the age differences are small, and often less than the SED time step of 1 Myr. We compare the widths of these distributions to the individual age uncertainties at the end of this section.

To test whether the age distributions for the three H$\alpha$ morphological classes are significantly different, we perform Kolmogorov-Smirnov (KS) tests in order to determine the probability that each class of objects originates from the same parent distribution, with the results summarized in Table~\ref{tab:table4}.  We perform the test for both isolated and non-isolated cluster samples, as well as the two samples combined, to improve the overall number statistics.

\begin{table*}
\begin{tabular}{lllllll}
                & \multicolumn{2}{c}{Isolated}                & \multicolumn{2}{c}{Not Isolated} & \multicolumn{2}{c}{Total Sample}                      \\
\hline
H$\alpha$ Class & Padova                    & Geneva                    & Padova                       & Geneva      & Padova                    & Geneva                    \\
\hline
1 vs. 2         & 0.967                     & 0.889                     & 0.039                        & 0.036       & 0.020                  & 0.028                     \\
1 vs. 3         & < $p_{5\sigma}$ & < $p_{5\sigma}$ & < $p_{5\sigma}$    & 3.00E-04    & < $p_{5\sigma}$ & < $p_{5\sigma}$ \\
2 vs. 3         & 0.024                     & 1.00E-05                  & 0.097                        & 0.006       & 5.00E-05                  & < $p_{5\sigma}$ \\
\hline
\end{tabular}
\caption{KS test results comparing the age distributions of each H$\alpha$ morphological class (1 = concentrated, 2 = partially exposed, 3 = no emission). p-values represent the corresponding probabilities that two samples share the same parent distribution; $p_{5\sigma}$ represents a p-value that corresponds to a confidence level greater than 5$\sigma$.}
\label{tab:table4}
\end{table*}


Overall, when the cluster population is divided into isolated and non-isolated samples, we cannot reject the hypothesis that the age distributions for clusters with partially exposed H$\alpha$ morphologies have been drawn from the same parent sample as clusters with concentrated or absent H$\alpha$. As would be expected however, the clusters with no H$\alpha$ emission are found to be statistically different from those with concentrated H$\alpha$.  When the isolated and non-isolated samples are combined to increase the sample sizes, the statistical differences between the age distributions of the concentrated and partially exposed classes versus the distribution for clusters without H$\alpha$ are significant at $\geq$ 3$\sigma$ confidence, while the p-values remain at the $\sim$2$\sigma$ confidence level between the age distributions for clusters with concentrated and partially exposed H$\alpha$ morphologies. These results hold regardless of the adopted stellar extinction model, except the confidence level drops to $\sim$1$\sigma$ in distinguishing the distributions of clusters with concentrated and partially exposed H$\alpha$ morphologies when the starburst or differential-starburst extinction models are used.

\begin{figure}
\includegraphics[width =0.5\textwidth]{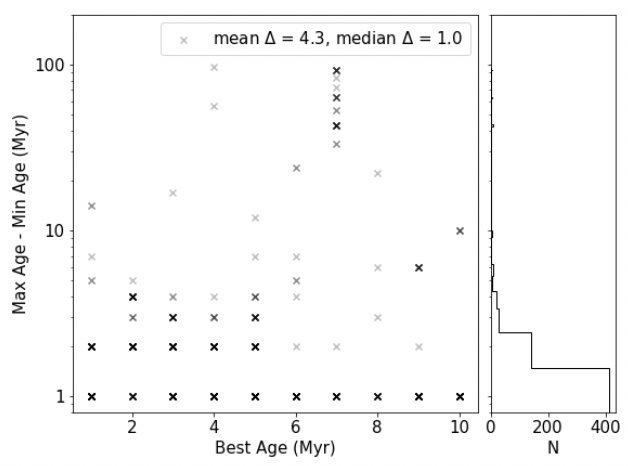}
\caption{Distribution of errors in Padova-based cluster age for all 654 clusters, where the errors are the differences between each cluster's minimum and maximum age. Darkness represents a greater density of points.}
\label{fig:Figure4}
\end{figure}

To determine if the width of the distributions represent real variations in the cluster ages of each morphological class, we must compare the widths to the errors. Fig.~\ref{fig:Figure4} displays the errors in cluster age, which is shown as the difference between the maximum and minimum age, plotted against the best estimates. While we see the mean error is buoyed up to $\sim$4 Myr due to extreme outliers, particularly due to the start of the red supergiant loop at 7 Myr in the SED models, the median error in age (1 Myr) is smaller than 10 of the 12 distributions' standard deviations, indicating that the errors are likely insufficient in describing the distribution widths. These results will be further discussed in Section ~\ref{sec:section61}.

\subsection{Reddening Statistics by Morphological Class}
\label{sec:section42}

\begin{figure*}
\includegraphics[width =1.0\textwidth]{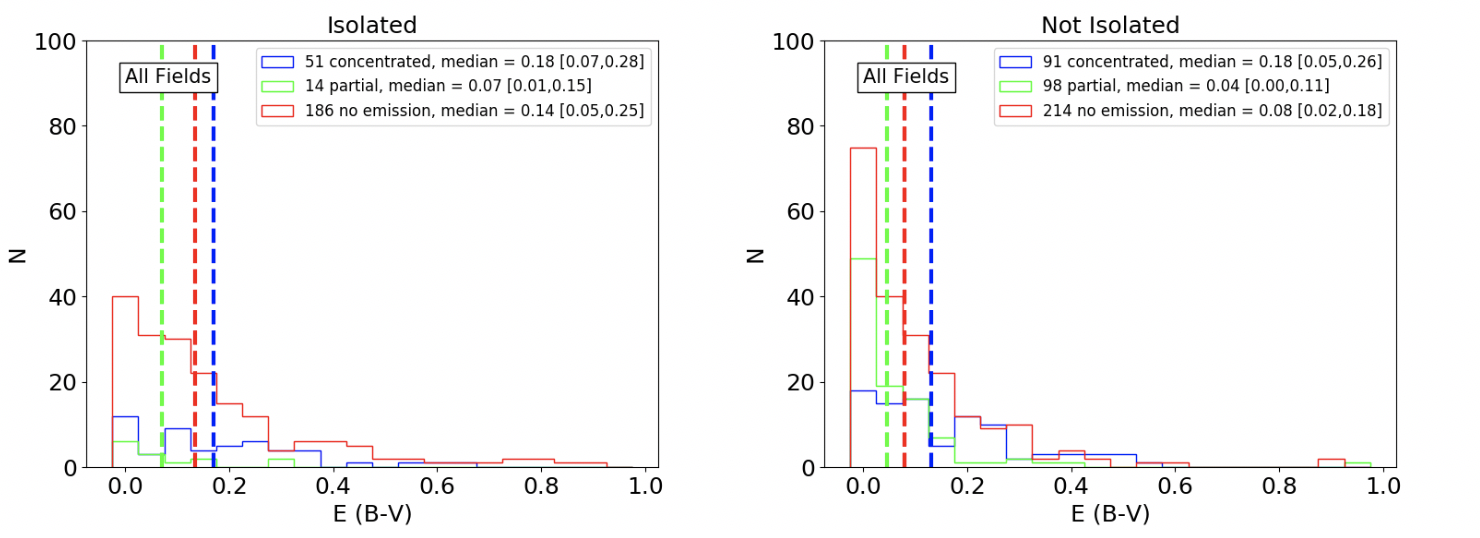}
\centering
\caption{E(B-V) histograms for all clusters across all the fields in the sample, using the Padova stellar evolutionary model. The left plot shows the 251 isolated clusters in the sample while the right plot shows the 403 clusters with neighboring clusters within 75 pc. The solid blue, green, and red lines represent clusters with H$\alpha$ emission that is concentrated, partially exposed, and absent, respectively. The number of clusters as well as the median reddening [first quartile, third quartile] for each of the three classes is displayed in the legend. Vertical dashed lines represent the medians for each class.}
\label{fig:Figure5}
\end{figure*}

The second parameter we examine is the reddening of the clusters in each morphological class, as a measure of dust in the region. Fig.~\ref{fig:Figure5} shows reddening histograms based on E(B-V) values derived from 5-band SED fitting for clusters across all six fields assuming Milky Way extinction \citep{CARDELLI89}, and divided into isolated (left) and non-isolated samples (right).

Interestingly, both isolated and non-isolated samples show that the clusters with partially exposed HII regions have the smallest median reddening, while the isolated concentrated and isolated no-emission H$\alpha$ morphologies have higher cluster reddening values. Regardless of the isolation category, stellar population model, extinction model, or using the mean or median values, we find consistent results.

Upon examination of the KS-test results of cluster reddening, the distributions are not found to be statistically different when the clusters are divided into isolated and non-isolated samples. After combining the samples, however, the KS tests show that all distributions are significantly different (>3$\sigma$) except for clusters with concentrated and no-emission H$\alpha$ morphologies, which are different at a $\sim$2$\sigma$ level, and these results are consistent across all stellar evolution and extinction models. The results of the KS tests are listed in Table~\ref{tab:table5}.

\begin{table*}
\begin{tabular}{lllllll}
                & \multicolumn{2}{c}{Isolated}                & \multicolumn{2}{c}{Not Isolated} & \multicolumn{2}{c}{Total Sample}                      \\
\hline
H$\alpha$ Class & Padova                    & Geneva                    & Padova                       & Geneva      & Padova                    & Geneva                    \\
\hline
1 vs. 2         & 0.107                     & 0.006                     & 7.00E-06                     & 4.00E-06    & < $p_{5\sigma}$ & < $p_{5\sigma}$ \\
1 vs. 3         & 0.665                     & 0.092                     & 0.016                        & 0.089       & 0.016                     & 0.021                     \\
2 vs. 3         & 0.290                     & 0.014                     & 0.013                        & 0.002       & 9.00E-05                  & 1.00E-06 \\
\hline
\end{tabular}
\caption{KS test results comparing the reddening distributions of each H$\alpha$ morphological class (1 = concentrated, 2 = partially exposed, 3 = no emission). p-values represent the corresponding probabilities that two samples share the same parent distribution.}
\label{tab:table5}
\end{table*}

\begin{figure}
\includegraphics[width =0.5\textwidth]{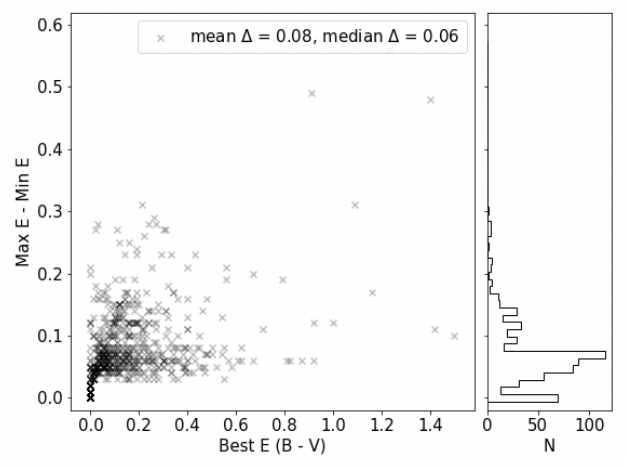}
\caption{Distribution of errors in Padova-based cluster reddening for all 654 clusters, where the errors are the differences between each cluster's minimum and maximum reddening. Darkness represents a greater density of points.}
\label{fig:Figure6}
\end{figure}

Fig.~\ref{fig:Figure6} displays the error in cluster reddening, which is shown as the difference between the maximum and minimum reddening, plotted against the best estimates for each cluster. We find that both the mean error (0.08) and median error (0.06), are smaller than the standard deviation of every distribution (ranging from 0.11 to 0.22), regardless of isolation category or stellar population model, again indicating that the error is insufficient in describing the width of the distributions. We will discuss these results in Section~\ref{sec:section63}.

\section{Stochasticity}
\label{sec:section5}

Our sample is dominated by low mass clusters.  The median mass of our entire cluster sample is 1100 $M_\odot$, with over 90$\%$ being less massive than 5000 $M_\odot$ (see also Table~\ref{tab:table3}), the limit used by \citet{HOLLYHEAD15} to minimize the impacts of stochastic (i.e. random) sampling in their analysis of clusters in M83.  At such masses, the initial mass function is not fully sampled and the relationship between physical and photometric properties is non-deterministic, which leads to a broad posterior probability distribution function (PDF) that is not well-characterized by a single best fit (e.g., \citealt{FOUESNEAU10,KRUMHOLZ15}).  Here, we look for evidence of this stochasticity and explore methods to mitigate against its impact.

\subsection{Color-Color Analysis}

Color-color plots are useful tools in examining the quality of SED age estimates relative to a model as well as highlighting any potential outlier clusters. As examined by \citet{FOUESNEAU12}, the predicted optical fluxes of a Monte Carlo sample of star clusters are spread quite widely in color-color space, especially for clusters below 10$^{4}$ $M_\odot$, where 96$\%$ of our sample lies. This is due to the fact that the integrated flux of such lower-mass clusters can be greatly influenced by the presence or absence of a few bright stars introduced by stochastic sampling of the IMF.

\begin{figure*}
\includegraphics[width =0.75\textwidth]{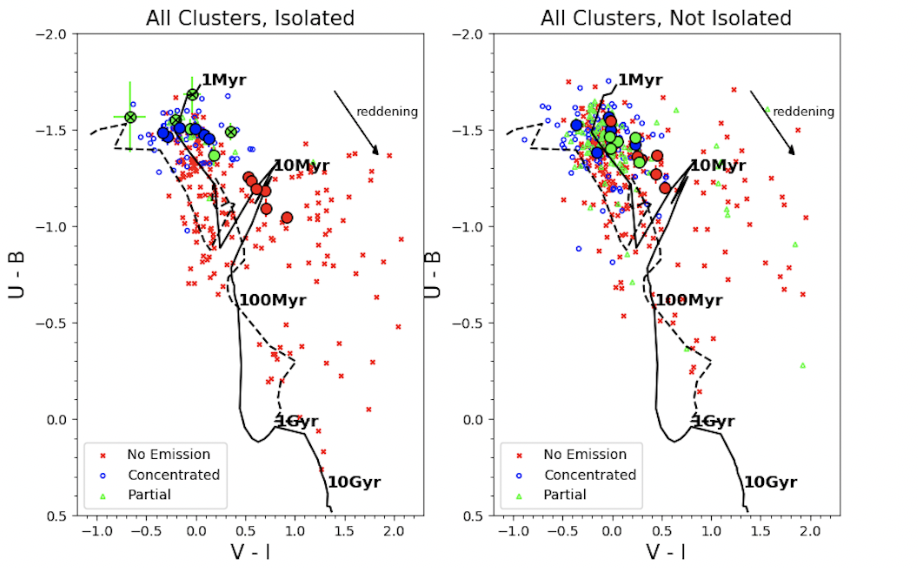}
\centering
\caption{(U-B) vs. (V-I) diagrams for all 654 clusters across all six fields. U, B, V, and I represent the F336W, F435W (NGC 1313) or F438W (NGC 4395, NGC 7793), F555W, and F814W filters, respectively. The 251 isolated clusters are shown in the left plot while the 403 non-isolated clusters are plotted on the right. Z = 0.02 (solid line) and Z = 0.004 (dashed line) Yggdrasil model tracks are overplotted (\citet{ZACKRISSON11}). Blue circles, green triangles, and red Xs represent clusters with H$\alpha$ morphologies classified as concentrated, partially exposed, and no-emission, respectively. An $A_V$ = 1.0 reddening vector is displayed in the upper-right corner of each plot. Large, black-outlined circular points represent the composite clusters from each field, provided here as an illustrative reference (see Section~\ref{sec:section53}).}
\label{fig:Figure7}
\end{figure*}

Fig.~\ref{fig:Figure7} shows observed (U-B) vs. (V-I) plots for all 654 young clusters across all six fields, split into isolated and non-isolated samples. There are a few notable features that are immediately apparent in these plots. As expected from the median age of each morphological class, we find clusters with no H$\alpha$ emission at the older end of the model predictions (red Xs), while the clusters with concentrated (blue circles) and partially exposed (green triangles) H$\alpha$ classes roughly overlap at the youngest end. In addition, while each class displays some spread in color space, we see that clusters with no H$\alpha$ emission have the largest degree of color spread. While the degree of spread may be surprising given that the clusters were selected to be younger than 10 Myr, it is qualitatively similar to the spread found in the stochastic sampling model produced by \citet[Figure 2]{FOUESNEAU12}.

\subsection{Influence of Red Supergiants}
\label{sec:section52}

\begin{figure}
\includegraphics[width =0.5\textwidth]{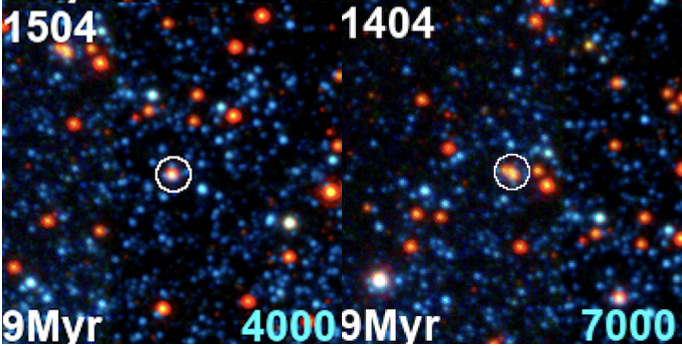}
\centering
\caption{RGB images of isolated clusters with no H$\alpha$ emission which appears red in V-I due to the presence of one or more bright red sources within the aperture radius. These stamps have been modified to highlight the red sources (B = B, G = V, R = I) instead of (absent) H$\alpha$.}
\label{fig:Figure8}
\end{figure}

The clusters in our sample without H$\alpha$ emission display a similar spread in (U-B) vs. (V-I) space to that observed by \citet{FOUESNEAU12}. This prompts us to re-inspect the postage stamps of each cluster in order to gain insight into the color-spread. Fig.~\ref{fig:Figure8} displays the postage stamps of two of the redder clusters with no H$\alpha$ emission.

We find that the flux from clusters without H$\alpha$ emission is significantly affected by the presence of bright red sources, which are presumably red supergiants. Of all the isolated clusters without H$\alpha$ emission, we find that each cluster located to the lower-right of the 10 Myr point on the model (V-I > 0.5, U-B < -0.5) was found to have at least one of these red sources within the photometric aperture. Overall, of all 400 clusters without H$\alpha$ emission in our sample, 134 (33$\%$) contained one or more bright red point source. 

\begin{figure*}
\includegraphics[width =1.0\textwidth]{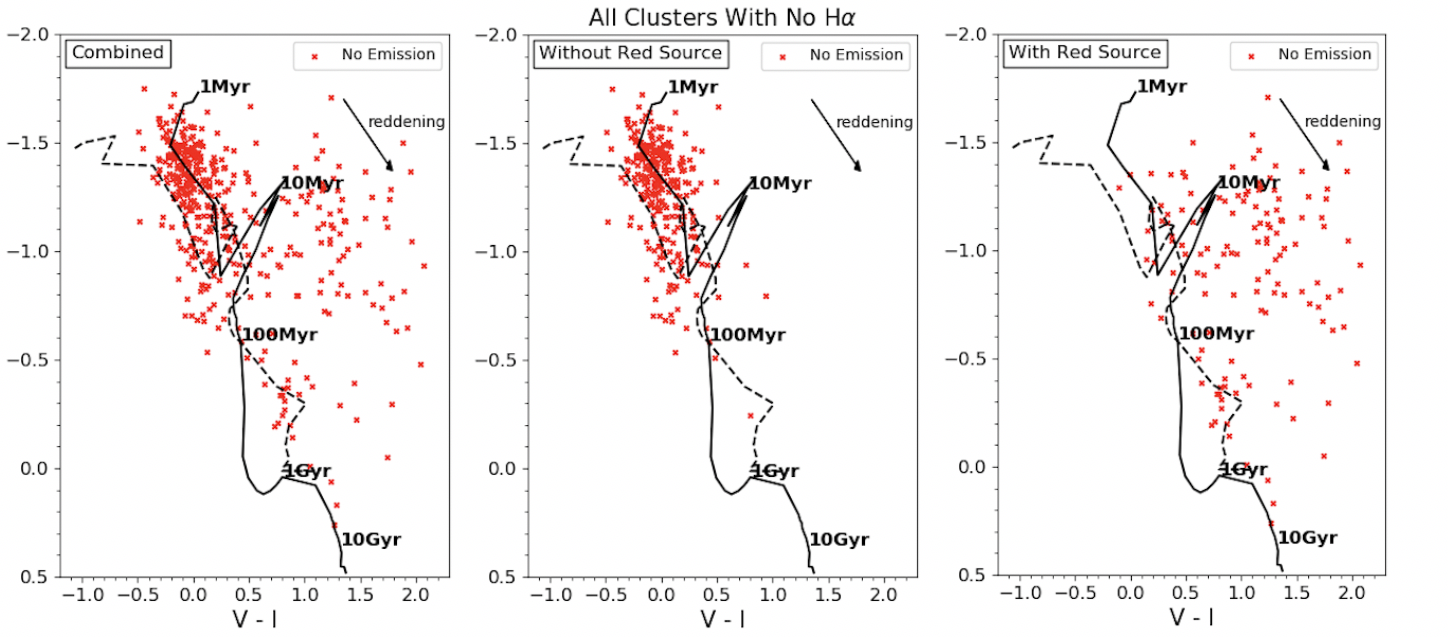}
\centering
\caption{(U-B) vs. (V-I) diagrams for all clusters without H$\alpha$ emission across all six fields. The left plot displays all 400 clusters, the middle plot displays the 266 clusters containing no red source, and the right plot displays the 134 clusters containing a red source.}
\label{fig:Figure9}
\end{figure*}

Fig.~\ref{fig:Figure9} shows color-color diagrams for all clusters without H$\alpha$ emission in our sample. We see that those without a red source are much less spread out and have a loci much closer to the model track. Those with a red source display a large spread toward the reddest (V-I) colors, consistent with the expectation that the integrated flux of lower-mass clusters is greatly affected by the presence (or absence) of a few individual bright sources.

\begin{figure}
\includegraphics[width =0.5\textwidth]{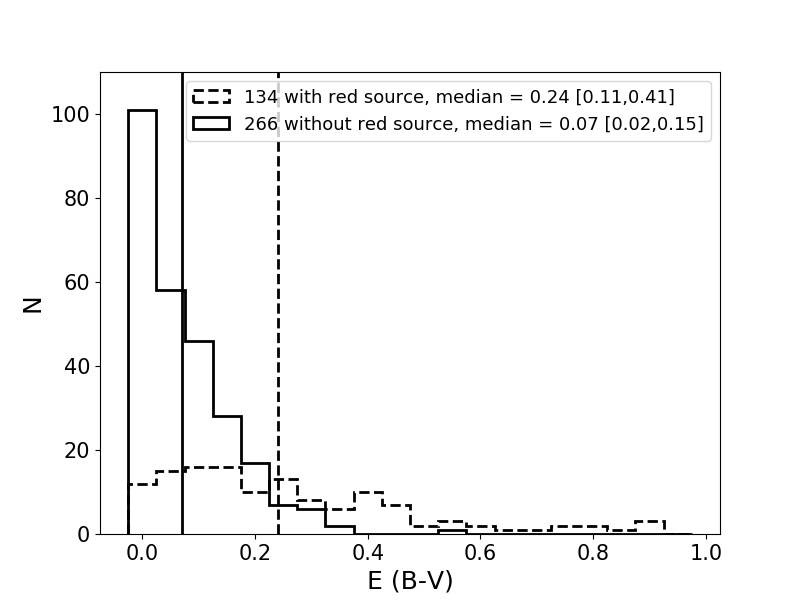}
\centering
\caption{E(B-V) histograms of clusters without H$\alpha$ emission across all six fields. The dashed line represents the 134 clusters which were found to contain a red point source while the solid line represents the 266 clusters not found to contain a red source. The median reddening of each category is provided in the legend and is plotted with vertical lines.}
\label{fig:Figure10}
\end{figure}

We also find that these clusters containing one or more bright red source are correlated with higher cluster reddening. Fig.~\ref{fig:Figure10} shows the E(B-V) distributions based on our reference catalog for all 400 clusters without H$\alpha$ emission, separately displaying the samples with and without bright red sources. We find that the 134 clusters containing a red source have an apparent median reddening value (0.24) more than three times that of the 266 clusters without one (0.07) and over 30$\%$ greater than the median reddening value of isolated clusters with concentrated H$\alpha$ ($\sim$0.18). Analysis of clusters in NGC 4449 by Whitmore et al. (in prep.) similarly revealed a significant fraction of clusters whose red colors are due to the presence of a dominating red source.

Within the no H$\alpha$ emission class, applying a KS test to the reddening distributions of clusters with a red source versus clusters without a red source reveals a confidence level greater than 5$\sigma$ that the two are not drawn from the same parent distribution. 

\subsection{Mitigating Stochastic Sampling Effects}
\label{sec:section53}

\begin{figure*}
\includegraphics[width =0.75\textwidth]{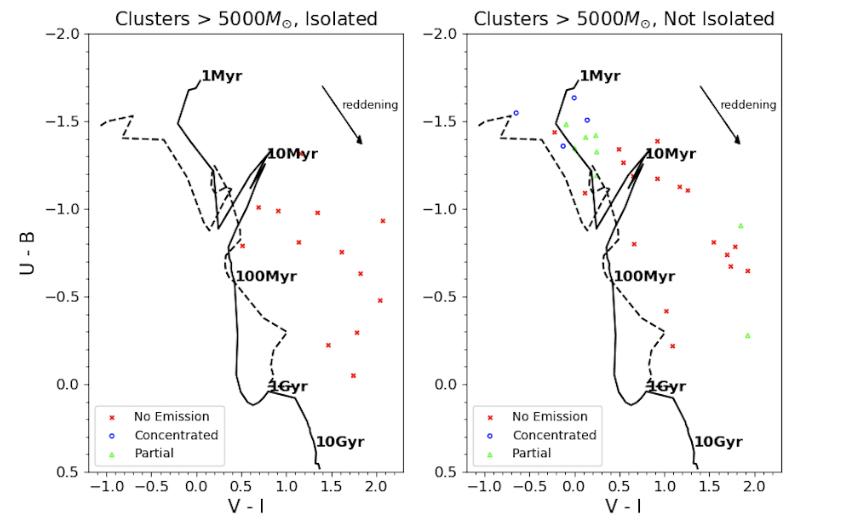}
\centering
\caption{(U-B) vs. (V-I) diagrams for all 42 clusters above the stochastic limit across all six fields. The 13 isolated clusters greater than 5000 $M_\odot$ are plotted on the left and the 29 non-isolated clusters greater than 5000 $M_\odot$ are plotted on the right. Blue circles, green triangles, and red Xs represent clusters with H$\alpha$ morphologies classified as concentrated, partially exposed, and no-emission, respectively.}
\label{fig:Figure11}
\end{figure*}

The uncertainty in the determination of physical properties of low mass clusters due to stochastic sampling is a particular challenge for our study. To provide better constraints on cluster properties despite the low masses of our cluster sample, we examine two potential methods of minimizing stochastic sampling effects. First, we simply limit the analysis to clusters above the threshold mass used by \citet{HOLLYHEAD15}, i.e. $\geq$ 5000 $M_\odot$. Fig.~\ref{fig:Figure11} shows color-color plots of all verified clusters with masses $\geq$ 5000 $M_\odot$, again divided by morphological class and isolation. 

While simply selecting only clusters with masses $\geq$ 5000 $M_\odot$ results in a very small sample (13 isolated, 29 non-isolated clusters), the little data remaining is consistent with what we found for all clusters (Fig.~\ref{fig:Figure7}), and still displays a significant spread in (U-B) vs. (V-I) space. Thus even when limiting to these higher mass clusters, stochasticity may still significantly impact the observed cluster properties: clusters with best fit ages $\leq$ 10 Myr can still be quite red and, in fact, all 13 of the isolated clusters above 5000 $M_\odot$ are amongst those found to contain one or more bright red source. 

In our second strategy to address the impact of stochasticity, we stack the fluxes of all individual clusters according to their H$\alpha$ morphologies (i.e. isolated/non-isolated and concentrated/partially exposed/no-emission). Here we create thirty-six composite clusters - one for each morphological class, isolation type, and within each field (6 bins x 6 fields). We note that NGC 4395N does not have non-isolated clusters with concentrated or partially exposed H$\alpha$ morphologies, and three other bins contain only a single cluster (see Table~\ref{tab:table6}) leaving us with 31 composite clusters. We sum the SED-determined masses of each of these composite clusters as an initial check on the composite mass. The summed masses range from 2.7x10$^{3}$ $M_\odot$ to 1.8x10$^{5}$ $M_\odot$ (see Table~\ref{tab:table6}).  For reference, 26 are above the 5000 $M_\odot$ limit used by Hollyhead et al. (2015).

Photometry for each of the NUV, U, B, V, and I bands are taken from the reference LEGUS catalog and converted into flux, utilizing the known zero points for each of the 8 \textit{HST} filters: F = 10$^{-0.4(m-m_z)}$, where $m_z$ is the zero-point magnitude. The cluster fluxes are summed in each of the 31 bins to produce the flux in the composite cluster. Photometric errors for composite clusters which are calculated by adding the constituent cluster errors in quadrature result in poor SED fitting. However, by using the median error for each composite cluster, the $\chi^{2}$ values for the SED fitting are comparable to the individual cluster $\chi^{2}$ values: the median $\chi^{2}$ is 2.1 and 1.4 for composite and individual clusters, respectively. The results of the fitting performed with the median errors are presented in Table \ref{tab:table6} and provide the basis for our analysis.

\begin{figure*}
\includegraphics[width =0.85\textwidth]{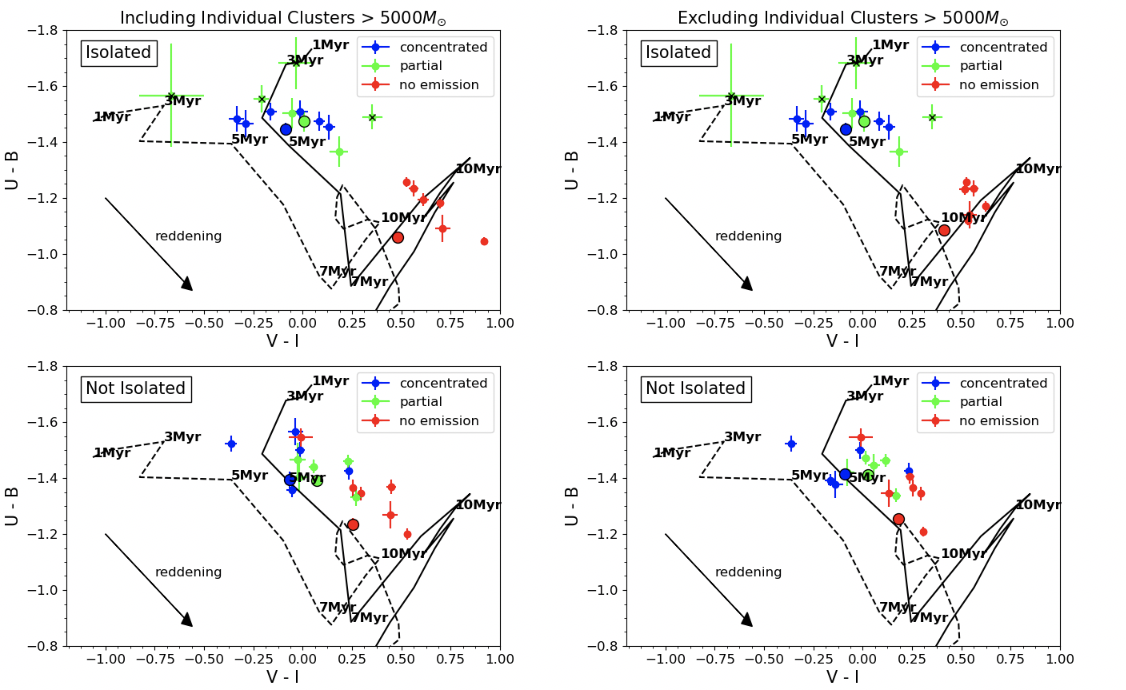}
\centering
\caption{(U-B) vs. (V-I) diagrams for all 31 composite clusters. The left and right columns display composite clusters, respectively, including and excluding individual clusters above 5000 $M_\odot$. The top row shows the 18 isolated composite clusters and the bottom row shows the 16 non-isolated composite clusters. Blue, green, and red points represent composite clusters with HII regions that are concentrated, partially exposed, and absent, respectively. Error bars represent the constituent cluster errors added in quadrature. The four composite clusters which have aggregate masses well below the 5000 $M_\odot$ limit are marked with a black X. The larger, black-outlined circles are provided for reference and represent the unweighted mean position of each morphological class across all six fields. It is important to note that these mean positions are not exact, as the V-band for NGC 4395 and NGC 7793 is covered by F438W, while NGC 1313 uses F435W.}
\label{fig:Figure12}
\end{figure*}

We plot the stacked fluxes of these composite clusters in (U-B) vs. (V-I) space to examine their new positions relative to the model tracks (Fig.~\ref{fig:Figure12}). Despite the significant percentage of individual clusters containing red sources (see Fig.~\ref{fig:Figure9} for their impact), the composite cluster sample is better behaved as they appear much closer to the model, likely due to the fact that stacking clusters diminishes the randomness in IMF sampling. The composite clusters with concentrated HII regions display some spread along the reddening vector while still being located at the youngest end of the model track. The composite isolated clusters without H$\alpha$ emission are located much closer to the model relative to their individual constituents, also displaying a much tighter locus found at the older end of the model, near the 10 Myr point. While the composite non-isolated clusters without H$\alpha$ also display a tighter locus than their constituent clusters, they generally appear closer to the younger end of the model than the isolated sample. The clusters with partially exposed morphologies suffer the most from small-number statistics, but ignoring the composite clusters which were made up of so few clusters that their aggregate masses remained well below the $M_\odot$ threshold, the rest still roughly overlap with clusters with concentrated H$\alpha$, if not lie between the other two classes.

Whether or not we include the few clusters above 5000 $M_\odot$ as part of these composite clusters did not significantly affect our overall conclusions. In (U-B) vs. (V-I) space, Fig.~\ref{fig:Figure12} displays composite clusters which include all individual clusters (Fig.~\ref{fig:Figure12}, left panels) with composite clusters which exclude the 42 individual clusters above 5000 $M_\odot$ (Fig.~\ref{fig:Figure12}, right panels). In this comparison, we find no significant color change in the concentrated and partially exposed H$\alpha$ classes. The greatest difference is in the clusters with no H$\alpha$ emission, where the lower-mass composite clusters appear slightly less red than their more-massive counterparts. Regardless of the inclusion of individual clusters above 5000 $M_\odot$, the evolutionary pattern remains consistent. 

\subsection{Measurement of the Physical Parameters for Composite Clusters}

We use SED fitting to calculate the age, mass, and reddening of each of our 31 composite clusters using the same method used to produce the physical cluster properties for the LEGUS catalogs \citep{ADAMO17}. The results of the SED fitting are shown for all 31 composite clusters in Table~\ref{tab:table6}, where the number of clusters making up each composite is also provided. Median age and reddening values are reported with their first and third quartiles while summed SED mass errors are calculated by adding all individual errors in quadrature. Errors in composite SED age, reddening, and mass each represent their individual SED error.

\begin{table*}
\begin{tabular}{lllllll}
\hline
\hline
         & \multicolumn{3}{c}{\bfseries Isolated, Concentrated}             & \multicolumn{3}{c}{\bfseries Non-isolated, Concentrated}         \\
         \hline
         & Number of & Median SED \textbf{Age}         & Composite SED \textbf{Age} & Number of & Median SED \textbf{Age}         & Composite SED \textbf{Age} \\
Field    & Clusters  & {[}25\%, 75\%{]} (Myr) & $\pm$ Error (Myr) & Clusters  & {[}25\%, 75\%{]} (Myr) & $\pm$ Error (Myr) \\
\hline
NGC1313W & 5         & 3.0 {[}3.0, 4.0{]}     & 4.0 $\pm$ 0.5         & 25        & 3.5 {[}3.0, 4.0{]}     & 3.0 $\pm$ 0.5         \\
NGC1313E & 10        & 4.0 {[}4.0, 4.0{]}     & 4.0 $\pm$ 0.5         & 15        & 4.0 {[}2.0, 4.0{]}     & 1.0 $\pm$ 1.0         \\
NGC4395N & 6         & 5.0 {[}3.5, 5.0{]}     & 5.0 $\pm$ 0.5         & 0         & -                      & -                 \\
NGC4395S & 9         & 5.0 {[}3.0, 5.0{]}     & 5.0 $\pm$ 0.5         & 16        & 3.0 {[}3.0, 5.0{]}     & 3.0 $\pm$ 0.5         \\
NGC7793W & 10        & 3.0 {[}1.3, 4.0{]}     & 2.0 $\pm$ 1.0         & 20        & 3.0 {[}2.0, 3.0{]}     & 2.0 $\pm$ 0.5         \\
NGC7793E & 11        & 2.0 {[}1.0, 2.0{]}     & 2.0 $\pm$ 0.5         & 15        & 1.0 {[}1.0, 2.0{]}     & 1.0 $\pm$ 0.5         \\
\hline
\hline
         & \multicolumn{3}{c}{\bfseries Isolated, Partially Exposed}        & \multicolumn{3}{c}{\bfseries Non-isolated, Partially Exposed}    \\
         \hline
         & Number of & Median SED \textbf{Age}         & Composite SED \textbf{Age} & Number of & Median SED \textbf{Age}         & Composite SED \textbf{Age} \\
Field    & Clusters  & {[}25\%, 75\%{]} (Myr) & $\pm$ Error (Myr) & Clusters  & {[}25\%, 75\%{]} (Myr) & $\pm$ Error (Myr) \\
\hline
NGC1313W & 1         & -                      & -                 & 20        & 4.5 {[}3.8, 7.0{]}     & 1.0 $\pm$ 0.5         \\
NGC1313E & 1         & -                      & -                 & 25        & 4.0 {[}3.0, 4.0{]}     & 3.0 $\pm$ 0.5         \\
NGC4395N & 1         & -                      & -                 & 0         & -                      & -                 \\
NGC4395S & 4         & 5.0 {[}5.0, 6.3{]}     & 15.0 $\pm$ 0.5        & 23        & 5.0 {[}5.0, 5.0{]}     & 15.0 $\pm$ 0.5         \\
NGC7793W & 3         & 2.0 {[}2.0, 3.0{]}     & 2.0 $\pm$ 0.5         & 19        & 3.0 {[}2.0, 3.0{]}     & 2.0 $\pm$ 0.5         \\
NGC7793E & 4         & 2.5 {[}1.8, 3.3{]}     & 2.0 $\pm$ 0.5         & 11        & 2.0 {[}2.0, 3.0{]}     & 2.0 $\pm$ 0.5         \\
\hline
\hline
         & \multicolumn{3}{c}{\bfseries Isolated, No Ha Emission}           & \multicolumn{3}{c}{\bfseries Non-isolated, No Ha Emission}       \\
         \hline
         & Number of & Median SED \textbf{Age}         & Composite SED \textbf{Age} & Number of & Median SED \textbf{Age}         & Composite SED \textbf{Age} \\
Field    & Clusters  & {[}25\%, 75\%{]} (Myr) & $\pm$ Error (Myr) & Clusters  & {[}25\%, 75\%{]} (Myr) & $\pm$ Error (Myr) \\
\hline
NGC1313W & 39        & 5.0 {[}4.0, 8.0{]}     & 15.0 $\pm$ 0.5        & 87        & 4.0 {[}3.0, 7.0{]}     & 13.0 $\pm$ 0.5        \\
NGC1313E & 29        & 6.0 {[}4.0, 7.0{]}     & 12.0 $\pm$ 0.5        & 31        & 4.0 {[}3.0, 4.0{]}     & 1.0 $\pm$ 0.5         \\
NGC4395N & 14        & 8.5 {[}5.3, 10.0{]}   & 10.0 $\pm$ 0.5        & 2         & 5.0 {[}5.0, 5.0{]}     & 5.0 $\pm$ 0.5         \\
NGC4395S & 20        & 6.0 {[}5.0, 10.0{]}    & 10.0 $\pm$ 0.5        & 40        & 5.0 {[}5.0, 6.0{]}     & 15.0 $\pm$ 0.5        \\
NGC7793W & 37        & 5.0 {[}2.0, 6.0{]}     & 6.0 $\pm$ 0.5         & 30        & 3.0 {[}3.0, 5.0{]}     & 2.0 $\pm$  0.5        \\
NGC7793E & 47        & 5.0 {[}4.0, 8.0{]}     & 20.0 $\pm$ 3.5        & 24        & 4.0 {[}2.0, 4.3{]}     & 1.0 $\pm$ 0.5 \\
\hline
\hline
\hline
         & \multicolumn{3}{c}{\bfseries Isolated, Concentrated}           & \multicolumn{3}{c}{\bfseries Non-isolated, Concentrated}       \\
         \hline
         & Number of & Median SED              & Composite SED  & Number of & Median SED              & Composite SED  \\
Field    & Clusters  & \textbf{E(B-V)} {[}25\%, 75\%{]} & \textbf{E(B-V)} $\pm$ Error & Clusters  & \textbf{E(B-V)} {[}25\%, 75\%{]} & \textbf{E(B-V)} $\pm$ Error \\
\hline
NGC1313W & 5         & 0.26 {[}0.11, 0.30{]}   & 0.15 $\pm$ 0.01    & 25        & 0.13 {[}0.09, 0.27{]}   & 0.23 $\pm$ 0.01    \\
NGC1313E & 10        & 0.22 {[}0.17, 0.26{]}   & 0.10 $\pm$ 0.01    & 15        & 0.22 {[}0.16, 0.30{]}   & 0.12 $\pm$ 0.03    \\
NGC4395N & 6         & 0.05 {[}0.02, 0.15{]}   & 0.13 $\pm$ 0.01    & 0         & -                       & -              \\
NGC4395S & 9         & 0.01 {[}0.00, 0.08{]}   & 0.06 $\pm$ 0.01    & 16        & 0.03 {[}0.00, 0.15{]}   & 0.02 $\pm$ 0.01    \\
NGC7793W & 10        & 0.28 {[}0.13, 0.38{]}   & 0.26 $\pm$ 0.02    & 20        & 0.11 {[}0.05, 0.17{]}   & 0.11 $\pm$ 0.01    \\
NGC7793E & 11        & 0.20 {[}0.12, 0.37{]}   & 0.22 $\pm$ 0.02    & 15        & 0.20 {[}0.10, 0.42{]}   & 0.22 $\pm$ 0.01    \\
\hline
\hline
         & \multicolumn{3}{c}{\bfseries Isolated, Partially Exposed}      & \multicolumn{3}{c}{\bfseries Non-isolated, Partially Exposed}  \\
         \hline
         & Number of & Median SED              & Composite SED  & Number of & Median SED              & Composite SED  \\
Field    & Clusters  & \textbf{E(B-V)} {[}25\%, 75\%{]} & \textbf{E(B-V)} $\pm$ Error & Clusters  & \textbf{E(B-V)} {[}25\%, 75\%{]} & \textbf{E(B-V)} $\pm$ Error \\
\hline
NGC1313W & 1         & -                       & -              & 20        & 0.06 {[}0.03, 0.12{]}   & 0.16 $\pm$ 0.02    \\
NGC1313E & 1         & -                       & -              & 25        & 0.09 {[}0.03, 0.13{]}   & 0.13 $\pm$ 0.01    \\
NGC4395N & 1         & -                       & -              & 0         & -                       & -              \\
NGC4395S & 4         & 0.00 {[}0.00, 0.04{]}   & 0.00 $\pm$ 0.01    & 23        & 0.00 {[}0.00, 0.00{]}   & 0.00 $\pm$ 0.01    \\
NGC7793W & 3         & 0.06 {[}0.04, 0.07{]}   & 0.07 $\pm$ 0.01    & 19        & 0.07 {[}0.00, 0.14{]}   & 0.13 $\pm$ 0.01    \\
NGC7793E & 4         & 0.14 {[}0.11, 0.22{]}   & 0.18 $\pm$ 0.02    & 11        & 0.06 {[}0.03, 0.13{]}   & 0.08 $\pm$ 0.02    \\
\hline
\hline
         & \multicolumn{3}{c}{\bfseries Isolated, No H$\alpha$ Emission}         & \multicolumn{3}{c}{\bfseries Non-isolated, No H$\alpha$ Emission}     \\
         \hline
         & Number of & Median SED              & Composite SED  & Number of & Median SED              & Composite SED  \\
Field    & Clusters  & \textbf{E(B-V)} {[}25\%, 75\%{]} & \textbf{E(B-V)} $\pm$ Error & Clusters  & \textbf{E(B-V)} {[}25\%, 75\%{]} & \textbf{E(B-V)} $\pm$ Error \\
\hline
NGC1313W & 39        & 0.26 {[}0.13, 0.46{]}   & 0.13 $\pm$ 0.01    & 87        & 0.14 {[}0.05, 0.21{]}   & 0.00 $\pm$ 0.01    \\
NGC1313E & 29        & 0.18 {[}0.08, 0.33{]}   & 0.06 $\pm$ 0.01    & 31        & 0.08 {[}0.04, 0.19{]}   & 0.27 $\pm$ 0.01    \\
NGC4395N & 14        & 0.08 {[}0.04, 0.14{]}   & 0.00 $\pm$ 0.01    & 2         & 0.01 {[}0.01, 0.02{]}   & 0.00 $\pm$ 0.01    \\
NGC4395S & 20        & 0.01 {[}0.00, 0.06{]}   & 0.00 $\pm$ 0.01    & 40        & 0.00 {[}0.00, 0.06{]}   & 0.00 $\pm$ 0.01    \\
NGC7793W & 37        & 0.10 {[}0.05, 0.17{]}   & 0.05 $\pm$ 0.01    & 30        & 0.10 {[}0.06, 0.14{]}   & 0.21 $\pm$ 0.01    \\
NGC7793E & 47        & 0.13 {[}0.05, 0.22{]}   & 0.00 $\pm$ 0.01    & 24        & 0.07 {[}0.04, 0.19{]}   & 0.16 $\pm$ 0.01   \\
\hline
\end{tabular}
\end{table*}

\begin{table*}
\begin{tabular}{lllllll}
\hline
\hline
         & \multicolumn{3}{c}{\bfseries Isolated, Concentrated}      & \multicolumn{3}{c}{\bfseries Non-isolated, Concentrated}      \\
         \hline
         & Number of    & Summed SED      & Composite SED  & Number of     & Summed SED       & Composite SED    \\
Field    & Clusters     & \textbf{Mass}  $\pm$  Error    & \textbf{Mass}  $\pm$  Error   & Clusters      & \textbf{Mass}  $\pm$  Error     & \textbf{Mass}  $\pm$  Error     \\
\hline
NGC1313W & 5            & 7.6 $\pm$ .4 E3       & 9.6 $\pm$ .3 E3      & 25            & 7.5 $\pm$ .3 E4        & 5.1 $\pm$ .1 E4        \\
NGC1313E & 10           & 1.2 $\pm$ .1 E4       & 1.0 $\pm$ .1 E4      & 15            & 1.0 $\pm$ .1 E5        & 7.7 $\pm$ 2.1 E4       \\
NGC4395N & 6            & 3.8 $\pm$ .3 E3       & 4.8 $\pm$ .2 E3      & 0             & -                & -                \\
NGC4395S & 9            & 3.9 $\pm$ .1 E3       & 4.1 $\pm$ .1 E3      & 16            & 8.0 $\pm$ .3 E3        & 6.0 $\pm$ .5 E3        \\
NGC7793W & 10           & 1.1 $\pm$ .1 E4       & 7.2 $\pm$ .1 E3      & 20            & 1.6 $\pm$ .1 E4        & 1.8 $\pm$ .2 E4        \\
NGC7793E & 11           & 1.9 $\pm$ .2 E4       & 1.7 $\pm$ .2 E4      & 15            & 2.5 $\pm$ .2 E4        & 2.2 $\pm$ .4 E4        \\
\hline
\hline
         & \multicolumn{3}{c}{\bfseries Isolated, Partially Exposed} & \multicolumn{3}{c}{\bfseries Non-isolated, Partially Exposed} \\
         \hline
         & Number of    & Summed SED      & Composite SED  & Number of     & Summed SED       & Composite SED    \\
Field    & Clusters     & \textbf{Mass}  $\pm$  Error    & \textbf{Mass}  $\pm$  Error   & Clusters      & \textbf{Mass}  $\pm$  Error     & \textbf{Mass}  $\pm$  Error     \\
\hline
NGC1313W & 1            & -               & -              & 20            & 6.0 $\pm$ .4 E4        & 7.2 $\pm$ .7 E4        \\
NGC1313E & 1            & -               & -              & 25            & 5.2 $\pm$ .3 E4        & 4.6 $\pm$ .1 E4        \\
NGC4395N & 1            & -               & -              & 0             & -                & -                \\
NGC4395S & 4            & 2.7 $\pm$ .3 E3       & 5.7 $\pm$ .1 E3      & 23            & 2.6 $\pm$ .1 E4       & 7.0 $\pm$ .1 E4        \\
NGC7793W & 3            & 5.1 $\pm$ .1 E3       & 8.3 $\pm$ .3 E3      & 19            & 2.3 $\pm$ .1 E4        & 2.6 $\pm$ .1 E4        \\
NGC7793E & 4            & 4.8 $\pm$ .4 E3       & 4.4 $\pm$ .2 E3      & 11            & 2.6 $\pm$ .4 E4        & 3.0 $\pm$ .4 E4        \\
\hline
\hline
         & \multicolumn{3}{c}{\bfseries Isolated, No H$\alpha$ Emission}    & \multicolumn{3}{c}\bfseries {Non-isolated, No H$\alpha$ Emission}    \\
         \hline
         & Number of    & Summed SED      & Composite SED  & Number of     & Summed SED       & Composite SED    \\
Field    & Clusters     & \textbf{Mass}  $\pm$  Error    & \textbf{Mass}  $\pm$  Error   & Clusters      & \textbf{Mass}  $\pm$  Error     & \textbf{Mass}  $\pm$  Error     \\
\hline
NGC1313W & 39           & 1.8 $\pm$ .2 E5       & 1.6 $\pm$ .1 E5      & 87            & 2.6 $\pm$ .2 E5        & 2.4 $\pm$ .1 E5        \\
NGC1313E & 29           & 8.8 $\pm$ 1.8 E4      & 9.5 $\pm$ .2 E4      & 31            & 1.1 $\pm$ .3 E5        & 1.4 $\pm$ .1 E5        \\
NGC4395N & 14           & 2.1 $\pm$ .1 E4       & 1.7 $\pm$ .1 E4      & 2             & 3.8 $\pm$ .3 E3        & 3.7 $\pm$ .1 E3        \\
NGC4395S & 20           & 3.4 $\pm$ .2 E4       & 3.4 $\pm$ .1 E4      & 40            & 1.0 $\pm$ .1 E5        & 1.6 $\pm$ .1 E5        \\
NGC7793W & 37           & 3.0 $\pm$ .6 E4       & 1.1 $\pm$ .1 E4      & 30            & 3.2 $\pm$ .4 E4        & 6.1 $\pm$ .1 E4        \\
NGC7793E & 47           & 5.9 $\pm$ .4 E4       & 7.3 $\pm$ 2.2 E4     & 24            & 2.8 $\pm$ .1 E4        & 3.8 $\pm$ .1 E4   \\ 
\hline
\end{tabular}
\caption{SED fitting results of all 31 composite clusters, separated into age results (top three tables), reddening results (middle three tables), and mass results (lower three tables). For each of the three metrics, the top table lists the six concentrated composite clusters, the middle table lists the six partially exposed composite clusters, and the bottom table lists the six no-emission composite clusters, each divided into isolated and non-isolated categories. The median age, median reddening, and sum total mass of individual clusters in each of the 31 composite clusters is provided adjacent to the corresponding composite SED result.}
\label{tab:table6}
\end{table*}

In examining the ages of the composite clusters, we first find a progression consistent with the analysis of the individual clusters: on average, the concentrated class comprises the youngest composite clusters (3.7 Myr isolated; 2.0 Myr non-isolated), the class without H$\alpha$ emission is the oldest (12.2 Myr isolated; 6.2 Myr non-isolated), and the partially exposed class lies between the two (6.3 Myr isolated; 4.6 Myr non-isolated), but the few data points we have overall show small differences between clusters with concentrated and partially exposed morphologies. It is also notable that 5 of the 6 composite clusters without H$\alpha$ emission have ages between 10 and 20 Myr, despite being comprised of \textit{apparently} young ($\leq$ 10 Myr) clusters. While the SED-fitting of composite clusters assumes a single stellar population age, we are in fact stacking clusters with a range of ages, particularly for the no-emission class. Regardless of this fact, we see an older age for composite isolated clusters without H$\alpha$ than the apparent median age of their constituents.

\begin{figure*}
\includegraphics[width =0.75\textwidth]{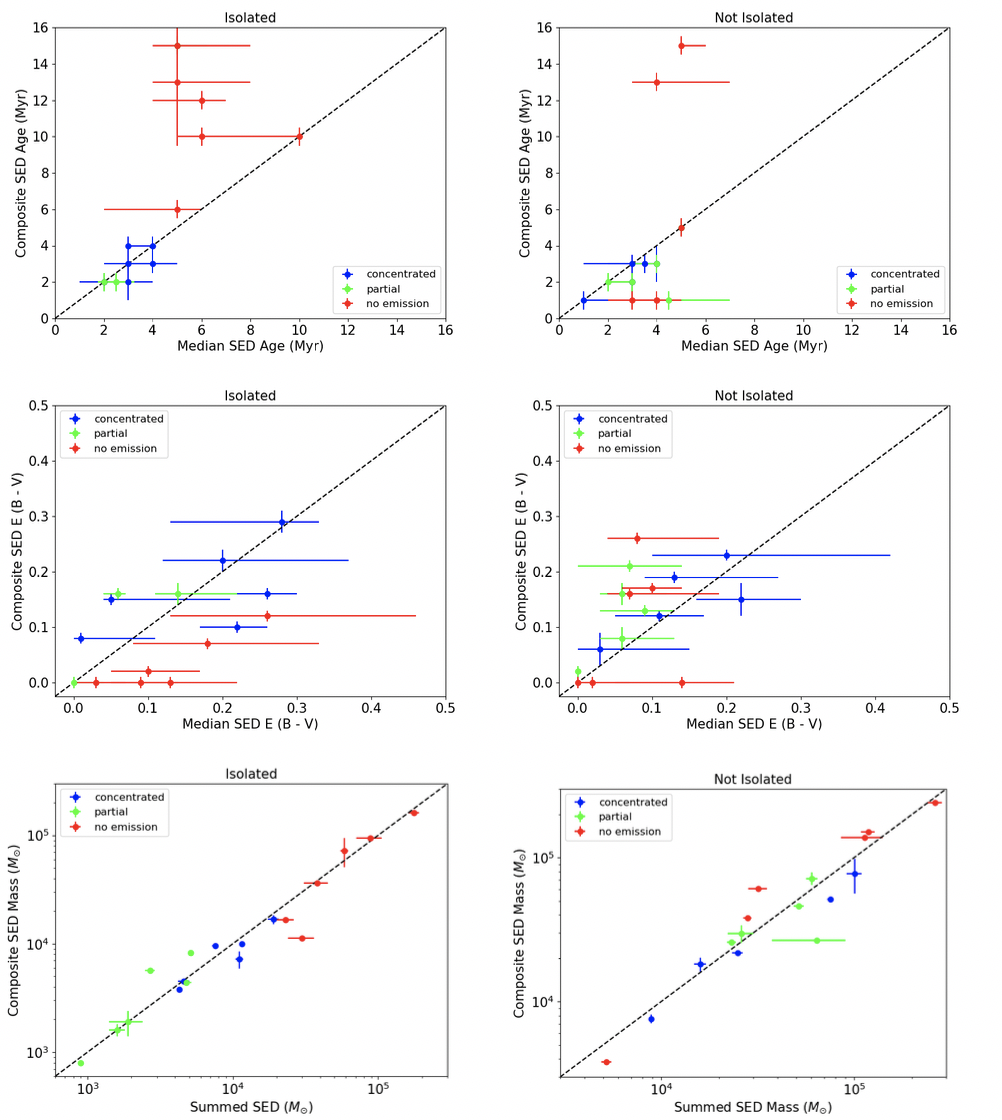}
\centering
\caption{Cluster age, reddening, and mass plots comparing the composite cluster SED results with the median constituent cluster SED values. Blue, green, and red points represent concentrated, partially exposed, and no-emission H$\alpha$ morphologies, respectively. Top row: SED age of composite clusters vs. Median SED age of their constituents. Middle row: SED reddening of composite clusters vs. Median SED reddening of their constituents. Bottom row: SED mass of composite clusters vs. Summed SED mass of their constituents. The left and right columns display the isolated and non-isolated samples, respectively. Horizontal error bars for age and reddening represent the 25$\%$ and 75$\%$ quartile values for constituent clusters; horizontal errors for mass depict the standard deviation in the distribution. Vertical error bars represent the SED error for the composite cluster.}
\label{fig:Figure13}
\end{figure*}

Fig.~\ref{fig:Figure13} shows plots comparing the SED-fit age, reddening, and mass of the composite clusters with the mean age, reddening, and mass of their constituent clusters. We find the masses determined from SED fitting for the composite clusters correlate well with their aggregate masses across all fields, classes, and isolation. There also seems to be a correlation between the age of composite isolated clusters versus the median values of their constituents for those with H$\alpha$ (concentrated and partially exposed classes), and we find similar agreement in those classes for reddening as well, albeit with wide distributions. Where we see the greatest discrepancy in cluster properties is in the ages and reddenings of clusters classified as no-emission. For the isolated sample, the mean and median ages of all individual clusters with no H$\alpha$ are $\sim$5 Myr, while the mean and median ages of the composite isolated clusters with no emission are $\sim$12 Myr, and all composite isolated clusters have best ages $\geq$ the mean or median age of their constituent clusters. Correspondingly, the median reddening of isolated clusters drops significantly from 0.14 for the constituent clusters to $\sim$0.03 for the composite clusters, and all isolated composite clusters have less reddening than the median of their constituents. 

The non-isolated sample of composite clusters without H$\alpha$, however, shows mixed results. While two of the composite clusters have significantly older ages and smaller reddening than the median values of their constituent clusters, as is the case for the isolated sample, three composite clusters show the opposite result, namely that they have younger ages ($\lesssim$ 2x; all 1-2 Myr) and correspondingly larger reddening (>2x) than the medians of their constituents. This is puzzling because, although they do appear closer to the younger end of the model than the isolated sample in (U-B) vs. (V-I) space (Fig.\ref{fig:Figure12}), the reddening vector would not appear to trace them back to the 1Myr point in the model but rather $\gtrsim$ 3Myr. The implications of these results will be further discussed in Section \ref{sec:section61}.

\section{Discussion}
\label{sec:section6}

\subsection{H\texorpdfstring{$\alpha$}{} Morphology Evolution and Timescales}
\label{sec:section61}

Examination of the age distributions of the individual star clusters as a function of H$\alpha$ morphological class (Section~\ref{sec:section41}) shows that although the distributions are broad and overlap each other, the mean and median ages provide evidence for a temporal progression in H$\alpha$ morphology. The star clusters with concentrated H$\alpha$ have the youngest average ages ($\sim$3 Myr), those with partially exposed morphology are older ($\sim$4 Myr), and those with no H$\alpha$ emission are the most evolved (>5Myr).  Consistent results are found when the sample is divided into clusters that are isolated, and those that have a neighbor (within 75 pc).

When KS tests are performed on isolated and non-isolated cluster sub-samples separately, we find that the null hypothesis that the age distributions for the clusters with concentrated and partially exposed H$\alpha$ morphologies are drawn from the same parent sample cannot be rejected with high certainty ($\sim$2$\sigma$), and the same is true for clusters without H$\alpha$ and those with partially exposed H$\alpha$ morphologies. When the samples are combined, however, KS tests reveal greater confidence levels in the uniqueness of all distributions, and while the confidence level is only just above 2$\sigma$ for the distributions of clusters with concentrated and partially exposed H$\alpha$ morphologies, which indicates a short clearing timescale ($\lesssim$ 1 Myr), each of the other distributions are found to be statistically different (>3$\sigma$). This is primarily a consequence of the decreased sizes of the subsamples (Table~\ref{tab:table2}).  Hence, the overall sample size must be increased to properly study the possible impact of nearby neighbors on HII region morphologies and gas clearing timescales.

Nevertheless, the age distributions for each of the H$\alpha$ morphological classes are wider than the formal uncertainties alone would allow. This likely indicates that H$\alpha$ morphology (and the gas removal process) depends on multiple parameters beyond the age of the parent star cluster, (e.g. nearby neighbors/local star formation density; dependence on metallicity; confining pressure of the ISM).

The age progression results just discussed are model dependent, and are based on the Padova models adopted for the LEGUS "reference" catalogs of star cluster properties, and using the Milky Way extinction model \citep{ADAMO17}.  When ages are instead derived from SED fitting to Geneva models or alternate extinction models (i.e. starburst, differential-starburst), the age distributions of the clusters with concentrated and partially exposed H$\alpha$ morphologies have consistent means and medians ($\sim$3-3.5 Myr).  KS tests confirm that the null hypothesis that these ages have been drawn from the same parent sample cannot be rejected with high certainty ($\lesssim 2\sigma$).  Of course, the statistical differences between the age distributions for those clusters with H$\alpha$ emission (whether concentrated or partially exposed) and those with no H$\alpha$ emission remain significant. 



Independent of the age distributions for each of the H$\alpha$ morphological classes, we can infer the lifetimes of the classes by examining their relative fractions, assuming that our sample is statistically representative of all clusters $\leq$ 10 Myr.  As shown in Table~\ref{tab:table7}, we see that the time a cluster spends in the concentrated H$\alpha$ stage is $\sim$2 Myr for the total sample as well as for the isolated and non-isolated sub-samples.  While there are greater discrepancies between the isolated and non-isolated samples for the partially exposed class, we can infer a range of timescales, $\sim$0.5-2.5 Myr, for this stage. This could be due to a number of environmental factors and perhaps is worthy of future investigation. The "length of time" spent in the no-emission class is then simply the remaining time left in the cluster's first 10 Myr, not the total length of time spent without H$\alpha$ emission.  For the total sample, the implied average ages for the clusters with concentrated H$\alpha$ and partially exposed morphologies are $\sim$1 Myr and $\sim$3 Myr, which are somewhat lower than the averages determined from analysis of the age distribution.

\begin{table*}
\begin{tabular}{lllllll}
\hline
                  & \multicolumn{2}{c}{Isolated} & \multicolumn{2}{c}{Non-Isolated} & \multicolumn{2}{c}{Combined} \\
                  \hline
H$\alpha$ Class   & \%      & Length of Stage    & \%        & Length of Stage      & \%      & Length of Stage    \\
\hline
Concentrated      & 20.5    & 2.1 Myr            & 22.5      & 2.3 Myr              & 21.7    & 2.2 Myr            \\
Partially Exposed & 5.5     & 0.6 Myr            & 24.7      & 2.5 Myr              & 17.3    & 1.7 Myr            \\
No-Emission       & 74      & 7.4 Myr            & 52.8      & 5.3 Myr              & 61      & 6.1 Myr \\
\hline
\end{tabular}
\caption{Relative fractions of, and inferred length of time spent in, each morphological class for the 251 isolated clusters, 403 non-isolated clusters, and the entire 654 cluster sample.}
\label{tab:table7}
\end{table*}

The characteristic ages of the H$\alpha$ morphological classes measured from SED fitting of the composite clusters (Table \ref{tab:table6}) yields a picture which is generally consistent with the analysis of the age distributions of the individual clusters, and the lifetimes inferred from the relative fractions of clusters in each class.  The composite clusters with concentrated H$\alpha$ morphologies show the youngest average ages ($\sim$3 Myr) and those where the cluster is partially exposed have slightly older ages ($\sim$5 Myr).  For the composites constructed from the isolated clusters with no H$\alpha$ emission, the ages are between 6 and 20 Myrs.  Interpretation of the non-isolated composite clusters without H$\alpha$ emission is less straightforward.  The positions of the non-isolated composite clusters on a color-color diagram (Fig. \ref{fig:Figure12}, bottom panels) show that those without H$\alpha$ emission are generally redder (and presumably older) than the composites with concentrated or partially exposed H$\alpha$ morphologies.  Yet, three of these composites have SED fit ages of 1-2 Myr, while the other three have ages between 5 and 15 Myr.  More work is needed to understand why the SED-fit ages for these composites are so low. If the low ages are robust, a speculation is that this might be due to the confounding effects cluster neighbors have on gas clearing (i.e., clusters that are young and still have ionizing OB stars may have had their gas pushed away by other nearby stellar populations).


Taken altogether, this analysis suggests that gas clearing begins early ($\lesssim3$ Myr) and occurs quickly ($\lesssim1$ Myr).  Such timescales provide evidence that young star clusters begin clearing their gaseous surroundings prior to the onset of the first SNe, due to radiation pressure and winds from their massive stars.  These findings are consistent with a range of previous results.  


In a similar analysis with HST H$\alpha$ images for 91 clusters in M83, \citet{HOLLYHEAD15} study found that clusters initially begin to remove gas at an age of $\sim$2Myr, and also found that the median cluster ages of their equivalent concentrated, partially exposed, and no-emission H$\alpha$ morphologies to be $\sim$4Myr, $\sim$5Myr, and $\sim$6Myr, respectively. 

\citet{GRASHA18}'s study of the LEGUS galaxy NGC 7793 examines clearing timescales by associating star clusters with their nearest GMC based on ALMA CO data. They determine the clearing timescale by tracking how the distribution in the age of the cluster populations changes as a function of their distance from the center of every GMC, and thus determine that clusters dissociate from their GMCs at ages of 2-3 Myr. 


\citet{KRUIJSSEN19} applied a statistical method to the combined observations of molecular gas and H$\alpha$ emission from young star clusters in NGC300, and subsequently applied the method to 9 other nearby spiral galaxies in Chevance et al. (in prep.), in order to characterize the correlation between GMCs and star formation. For NGC300, they found that GMCs and HII regions coexist on average for 1.5 $\pm$ 0.2 Myr while the larger sample showed coexistence timescales between 1-5Myr, both of which support the conclusion that feedback prior to the onset of supernovae, such as stellar winds and radiation pressure, plays an important role in the dispersal of a star cluster's natal cloud.
 
Simulations also support this scenario. \citet{KIM18} modeled the dispersal of GMCs by photoionization and radiation pressure and found a range of cloud destruction times between 2 and 10 Myr after the onset of radiation feedback. For their fiducial model (initial cloud radius = 20pc, mass = 10$^{5}$ $M_\odot$, $t_{ff}$ = 4.7 Myr), they found that $\sim$50$\%$ of their simulation box (80pc x 80pc) was filled with ionized gas within 0.8Myr after the first stars were formed, and had mostly cleared all dense gas within a 10pc radius of the cluster within 1 $t_{ff}$ (4.7 Myr). This was performed without the aging of the stellar populations and thus these timescales serve as lower limits for radiation feedback. However, this model also does not include SNe, which would produce their own feedback as early as $\sim$3Myr. 

\citet{TREMBLIN14} used 1D expansion models to investigate the development of HII regions around young star clusters using isochrone-based age estimates. By testing their models on the Rosette, M16, RCW79, and RCW36 HII regions (which would either have concentrated or partially exposed Halpha morphologies in this study), they found their dynamical ages to agree with photometric ages from previous results, and to support the early onset of gas clearing by ~2Myr. Furthermore, they found that for these four HII regions, a larger cloud radius correlated with an older age.  Correlations of age with HII region size has also reported by \cite{HOLLYHEAD15}, and \cite{WHITMORE11} and would be interesting to investigate in future work. 

It is interesting to note that when we examine the fields individually (see Table 6), we see that galaxies with higher metallicity (NGC 1313, NGC 7793; Z = 0.02) have younger median ages for clusters with concentrated and partially exposed H$\alpha$ morphologies ($\sim$2-4Myr) than those of the lower-metallicity galaxy (Z = 0.004), NGC 4395 ($\sim$5Myr).  A speculation is that longer clearing timescales are associated with lower metallicity systems, as winds would be weaker for lower metallicity stellar populations \citep{starburst99}.  Detailed studies of individual young clusters in metal-rich galaxies such as Westerlund 2 \citep{ZEIDLER2015} and NGC 3603, both of which are located in the Milky Way, \citep{PANG2011} have shown early signs of gas clearing. Images of these clusters (\citealt[Figure 2]{ZEIDLER2015}; \citealt[Figure 4]{PANG2011}) show a cavity formed around the central stars of at least $\sim$1-2pc (corresponding to our partially exposed class) while their estimated ages are $\sim$1-2 Myr. 

\subsection{Cluster Mass and Stochasticity}

\begin{figure}
\includegraphics[width =0.5\textwidth]{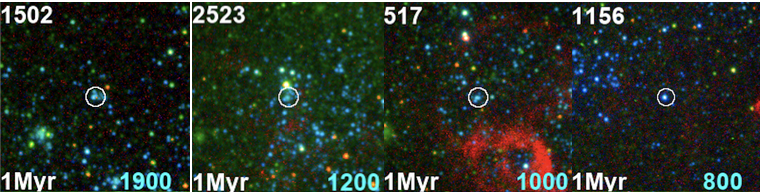}
\centering
\caption{Four 150pc x 150pc RGB postage stamps of the youngest (1 Myr) isolated clusters with HII regions classified as no-emission. Cluster ID is located in the upper left, age in the lower-left, and mass (in $M_\odot$) in the lower-right.}
\label{fig:Figure14}
\end{figure}

Since young clusters should have O-type stars ionizing their natal gas, it may be surprising that we find clusters that have estimated ages younger than 3 Myr (before the onset of SNe) with no immediate H$\alpha$ emission. Fig.~\ref{fig:Figure14} shows postage stamps of four of the youngest of these clusters classified as no-emission, each of which have SED-determined ages of 1 Myr.  There are at least three possible explanations for this phenomenon: 1) the SED determined ages for these clusters are incorrect, 2) these low-mass clusters are young but did not produce an O-type star, or 3) the density of HII is so low that the surface brightness of the H$\alpha$ emission is below our detection limit. 

\begin{figure}
\includegraphics[width =0.5\textwidth]{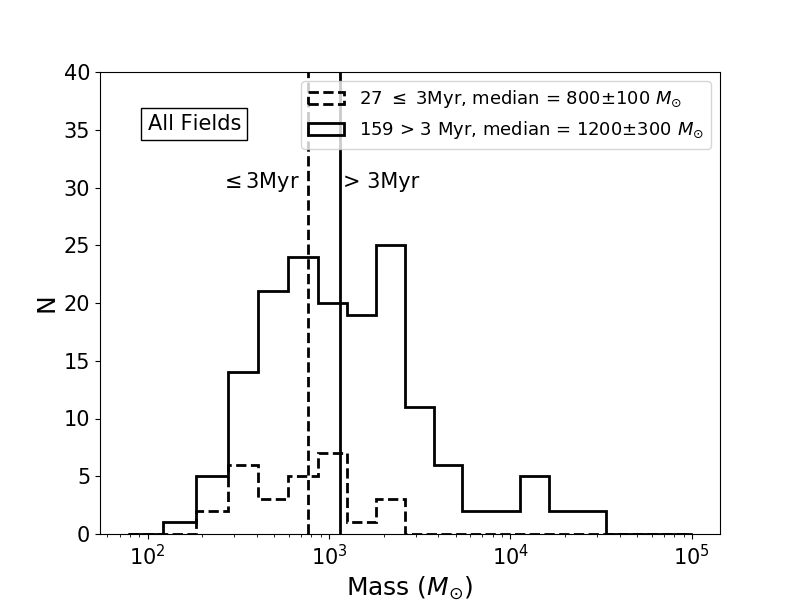}
\centering
\caption{Mass histogram of all 186 isolated clusters without H$\alpha$ emission. Clusters $\leq$ 3 Myr are plotted with a dashed line while clusters > 3 Myr are plotted with a solid line. The number of clusters as well as the median mass values for these two bins is provided in the legend and the medians are plotted with vertical lines.}
\label{fig:Figure15}
\end{figure}

 The mass distribution of clusters without H$\alpha$ emission may provide some insight. The mass histogram of two age bins in Fig.~\ref{fig:Figure15} shows that there is a small difference in median mass between those younger than 3 Myr (800 $M_\odot$) and those older than 3 Myr (1200 $M_\odot$). This could support the notion that the lower-mass clusters have a higher probability to not produce an O-type star to ionize its natal hydrogen gas. It is also plausible, however, that the low-mass nature of these clusters contribute to a longer dynamical timescale for the formation of each cluster and hence may add another confounding effect, as our models assume a single stellar population age for the entire cluster.

\begin{figure}
\includegraphics[width =0.5\textwidth]{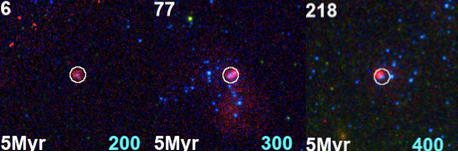}
\centering
\caption{150pc x 150pc RGB postage stamps of isolated clusters $\geq$ 5 Myr with concentrated HII regions. Cluster ID is located in the upper left, age in the lower-left, and mass (in $M_\odot$) in the lower-right.}
\label{fig:Figure16}
\end{figure}

We also find clusters with best-fit ages $\geq$ 5 Myr which show concentrated H$\alpha$ morphologies, after the typical 3-4 Myr timescale for SNe to begin clearing. 3 of the 6 isolated clusters with best ages $\geq$ 5 Myr and concentrated HII regions have their postage stamps shown in Fig.~\ref{fig:Figure16}. All of these older concentrated clusters were found in NGC 4395, the lowest-mass galaxy in our sample, and none were found in NGC 7793 and NGC 1313. The clusters themselves are relatively lower mass as well (from 200-500 $M_\odot$), thus stochastic effects may introduce larger uncertainties into the age determinations. 

\subsection{Cluster Reddening and Stochasticity}
\label{sec:section63}

Overall, in all three galaxies, we find relatively low reddening values for our clusters, where the range of our median E(B-V) values across the different morphological classes correspond to $A_V$ values between ~0.3 and ~0.6. \citet{KAHRE18}'s examination of extinction versus neutral H column density for NGC 7793 revealed similarly small extinction values, with the vast majority between $A_V$ values of 0.3 and 0.6, consistent with what we find in our cluster sample. 

A comparison of the color-color plots of our sample with the massive M83 clusters examined by \citet{WHITMORE11} and \citet{HOLLYHEAD15} show interesting differences. In the M83 studies, clusters with concentrated H$\alpha$ emission showed a high degree of reddening, with some clusters lying near the 10 Gyr point on the model, while the clusters with no H$\alpha$ emission had the bluest colors and the tightest locus. As cluster age increases, the concentration of gas, H$\alpha$ emission, and thus reddening are expected to decrease \citep{BARLOW93}. This is supported by the observations of \citet{HOLLYHEAD15}; namely that the clusters of M83 without H$\alpha$ emission were not nearly as spread out due to reddening as the clusters with concentrated H$\alpha$. In our sample, however, we see that clusters with no H$\alpha$ are the ones that have the reddest colors because a significant fraction of them contain a bright red source. 
Additionally, we find that clusters without H$\alpha$ have reddening values that are comparable to those which show concentrated H$\alpha$ morphologies.  This could be a result of overestimated reddening and corresponding underestimated ages, as also found by \citet{WHITMORE11} and Whitmore et al. (in prep.) in the analysis of low mass clusters without H$\alpha$ in M83 and NGC 4449.

Further for NGC 7793 we have compared the reddening values of these clusters based on SED fitting with extinction maps made from the Balmer decrement (H$\alpha$/H$\beta$) using VLT-MUSE (Multi Unit Spectroscopic Explorer) data (Della Bruna et al. in prep).  There are 65 clusters in our young cluster sample ($\leq$ 10Myr) which have MUSE coverage. For all 65 clusters, the median best SED-fit reddening is 0.11 while the median reddening determined with H$\alpha$/H$\beta$ via the MUSE data is 0.36. It is notable that only 7 of these 65 clusters have a best SED-fit E(B-V) greater than the Balmer calculated E(B-V) and each of those 7 clusters contain a bright red source. While clearly suffering from small number statistics, the data do seem to suggest that clusters containing a bright red source correlate with a larger SED-fit E(B-V) than the H$\alpha$/H$\beta$ determined E(B-V). 


The uniqueness of the reddening distributions of clusters with and without bright red sources (Fig.~\ref{fig:Figure10}), and comparison with the local gas extinction lead us to conclude that the presence of bright red sources produces a significant impact on the SED-fit reddening of a cluster. Thus we find that our sample of clusters with no H$\alpha$ emission have red colors, which are not necessarily due to dust, but rather because there is a bright red source within the aperture. 

Lastly, this likely overestimation of reddening for our clusters appears to be corrected by our stacking procedure. Compared to their constituent clusters, the older ages and smaller reddening values for our isolated composite clusters are more consistent with a cleared-out environment and thus potentially mitigates stochastic effects. As such, this procedure appears to provide a promising avenue for analyzing isolated clusters with stellar masses below the stochastic limit. More care must be taken when stacking non-isolated clusters, which show dichotomous results (see Section ~\ref{sec:section61} for this discussion).

\section{Summary and Future Work}
\label{sec:section7}

We examine $\sim$700 young ($\leq$ 10Myr) star clusters in the nearby spiral galaxies NGC 7793, NGC 1313, and NGC 4395 (d $\approx$ 4Mpc).  We study the H$\alpha$ morphology of the HII regions surrounding the clusters in relation to cluster age, reddening, and mass derived through SED-fitting of \textit{HST} NUV, U, B, V, I photometry.  The SED-fit properties are available via the LEGUS star cluster catalogs which have been publicly released through MAST.  The ultimate objective is to use the H$\alpha$ morphology around young star clusters to gain insight into the timescales, and thus the physical processes at work, in the clearing of a cluster's natal gas.  We classify the clusters in the sample according to: 1) visually-determined H$\alpha$ morphology (concentrated, partially exposed, and no-emission) and 2) whether they have neighboring clusters, which could affect the clearing timescales. A summary of our main results are as follows:

\begin{enumerate}
    
    \item The distributions of cluster ages for each of the H$\alpha$ morphological classes are consistent with the expected evolutionary sequence, as also found by studies of young star clusters in M83 (\citealt{WHITMORE11, HOLLYHEAD15}).  For the combined samples of isolated and non-isolated clusters, the median age of clusters (1) with concentrated H$\alpha$ is the lowest at $\sim$3 Myr; (2) which are partially exposed by their H$\alpha$ emission is $\sim$4 Myr; and (3) with no H$\alpha$ emission is the highest at > 5 Myr. The distinction between the ages of clusters with concentrated and partially exposed H$\alpha$ morphologies is model dependent, however. The mean and median ages of these classes based on the Geneva stellar evolution model (as opposed to our reference catalog, which uses the Padova model) or with alternate extinction models (i.e. starburst or differential-starburst extinction instead of Milky Way extinction) are consistent ($\sim$3-3.5Myr). Overall, this indicates that the clearing timescale is short -- on the order of or less than our SED time step of 1Myr.
    
    \item When the isolated and non-isolated samples are treated separately, KS tests cannot confirm with high certainty ($\sim$2$\sigma$) that the age distributions for all three morphological classes are statistically different. When the isolated and non-isolated samples are combined, however, the likelihood that they do not share a parent distribution is stronger ($\gtrsim$3$\sigma$). These results indicate that a larger sample size is needed to properly study the possible impact of nearby neighbors on HII region morphologies and gas clearing timescales.
    
    \item In contrast to previous studies and expectation, which find that clusters without H$\alpha$ emission are less reddened (e.g. \citealt{WHITMORE11,HOLLYHEAD15,BARLOW93}), we find comparable median E(B-V) values for clusters with concentrated H$\alpha$ and no H$\alpha$ emission (0.18 vs. 0.14, respectively).   However, the clusters with no H$\alpha$ tend to contain bright red point-like sources, and are significantly redder in the (U-B) vs. (V-I) diagram than clusters with concentrated H$\alpha$.  Given that these clusters have very low masses (several hundred $M_\odot$), we posit that this is the result of stochasticity in IMF sampling, and that the reddening has been overestimated for clusters containing bright red sources.
    
\end{enumerate}

Our experiments to mitigate stochastic effects by summing the fluxes of the clusters in each H$\alpha$ morphological class to synthesize more massive composite sources have yielded promising results.  From the observed properties of the composite clusters along with the ages, masses, and reddenings from SED fitting, we find that:

\begin{enumerate}

\item The colors of the composite clusters lie close to the evolutionary model track of a single-age population, and in general, have an age progression consistent with results based on the analysis of individual clusters.

\item The ages and reddening of composite clusters with concentrated and partially exposed morphologies are comparable to their constituents.

\item Isolated composite clusters with no H$\alpha$ are over twice as old ($\sim$12 Myr vs. $\sim$5 Myr) and have significantly less reddening ($\sim$5x) than their constituent clusters, which is more consistent with a cleared-out environment. The non-isolated sample, however, shows mixed results. While two of the composite clusters similarly have older ages and smaller reddening than their constituents, three composite clusters show the opposite result, namely having younger ages (all 1-2 Myr) and larger reddening (>2x) than their constituents. This is especially puzzling because of their position in (U-B) vs. (V-I) space: while they do appear closer to the younger end of the model than the isolated sample, the reddening vector would not appear to trace them back to the 1Myr point in the model but rather $\gtrsim$ 3Myr. We speculate that this could be the result of confounding effects cluster neighbors have on gas clearing.

\end{enumerate}
In this work, we have designed the analysis to rely on the LEGUS star cluster catalogs.  The star clusters, which are effectively single aged stellar populations, act as clocks which can be used to time the evolution of the HII region H$\alpha$ morphologies and clearing of the gas.  While this provides a statistically complete sample of star clusters for study, it does not allow us to answer questions about the fraction of the overall HII region population that are associated with star clusters, or to check whether the HII regions associated with the clusters studied here are a representative sample of the population.  Such questions require complete HII region catalogs to be developed (e.g., \citep{THILKER00, KRECKEL16}), and will be pursued in future work (e.g., Della Bruna et al., in prep).

Additionally, one of us (SH) visually classified the H$\alpha$ morphology of the clusters.  The usual drawbacks with human visual classification include the time consuming nature of the task, and the relative subjectivity of classification.  In the future, a more quantitative approach, which includes measurement of the concentration index of the H$\alpha$ emission, could be pursued.  A cursory examination of concentration indices of H$\alpha$ relative to the total H$\alpha$ flux for our sample confirms photometric differences between the morphological classes. Machine learning techniques could also be pursued using the classifications established here as the foundation of a training sample. Robust training of neural networks would require larger samples with human classifications than used here.
Such approaches could be used to classify the many thousands of additional young clusters in the full H$\alpha$ LEGUS galaxy sample, and facilitate future study of possible environmental dependences of HII region evolution timescales. 

\section*{acknowledgements}

Based on observations with the NASA/ESA/CSA Hubble Space Telescope which were retrieved from MAST at the Space Telescope Science Institute, operated by the Association of Universities for Research in Astronomy, Incorporated, under NASA contract NAS5-26555. The observations were obtained through \textit{HST} programs $\#$13364 and $\#$13773.  Support for these programs was provided through a grant from the STScI under NASA contract NAS5-26555.




\bibliographystyle{mnras}
\bibliography{sjh} 








\bsp	
\label{lastpage}
\end{document}